\documentclass[amssymb,usenatbib,useAMS,usedcolumns]{mn2e}
\usepackage{times} 
\usepackage{latexsym,amssymb,amsmath,amsfonts}
\usepackage{rotating} 
\usepackage{float}
\usepackage{graphicx}
\usepackage{natbib}
\usepackage{longtable}
\usepackage{url}
\usepackage{dcolumn}
\usepackage{textcomp}
\begin{document}
\newcommand{\hon}{\mbox{H\,{\sc i}}}
\newcommand{\oon}{\mbox{O\,{\sc i}}}
\newcommand{\cto}{\mbox{C\,{\sc ii}}}
\newcommand{\con}{\mbox{C\,{\sc i}}}
\newcommand{\sit}{\mbox{Si\,{\sc ii}}}
\newcommand{\sitr}{\mbox{Si\,{\sc iii}}}
\newcommand{\sto}{\mbox{S\,{\sc ii}}}
\newcommand{\non}{\mbox{N\,{\sc i}}}
\newcommand{\nto}{\mbox{N\,{\sc ii}}}
\newcommand{\mgt}{\mbox{Mg\,{\sc ii}}}
\newcommand{\mgo}{\mbox{Mg\,{\sc i}}}
\newcommand{\fet}{\mbox{Fe\,{\sc ii}}}
\newcommand{\fetr}{\mbox{Fe\,{\sc iii}}}
\newcommand{\pto}{\mbox{P\,{\sc ii}}}
\newcommand{\cat}{\mbox{Ca\,{\sc ii}}}
\newcommand{\nao}{\mbox{Na\,{\sc i}}}
\newcommand{\aro}{\mbox{Ar\,{\sc i}}}
\newcommand{\znt}{\mbox{Zn\,{\sc ii}}}
\newcommand{\mnt}{\mbox{Mn\,{\sc ii}}}
\def\h2{$\rm H_2$}
\def\Nh2{$N$(H${_2}$)}
\def\chin{$\chi^2_{\nu}$}
\def\chiu{$\chi_{\rm UV}$}
\def\sys{J0441$-$4313~}
\def\lya{\ensuremath{{\rm Ly}\alpha~}}
\def\lymana{\ensuremath{{\rm Lyman}-\alpha~}}
\def\kms{km\,s$^{-1}$}
\def\cms{cm$^{-2}$}
\def\cc{cm$^{-3}$}
\def\zabs{$z_{\rm abs~}$}
\def\zem{$z_{\rm em~}$}
\def\nhi{$N$($\hon$)~}
\def\ln{log~$N$}
\def\nh{$n_{\rm H}$}
\def\ne{$n_{e}$}
\def\21{21-cm}
\def\ts{T$_{s}$}
\def\th{T$_{01}$}
\def\ll{$\lambda\lambda$}
\def\l{$\lambda$}
\def\fc{$f_{c}$}
%
%=============================================================================================
%
\title[Cold parsec-scale gas]{
{%Understanding H$_{2}$ \& \21 absorption in the context of parsec-scale gas structure in a \zabs $\sim$ 0.1 sub-DLA 
Cold parsec-scale gas in a \zabs $\sim$ 0.1 sub-DLA with disparate \h2 and \21 absorption\thanks{Based 
on observations made with the NASA/ESA Hubble Space Telescope, obtained from the data archive at the 
Space Telescope Science Institute, which is operated by the Association of Universities for Research 
in Astronomy, Inc., under NASA contract NAS 5-26555; 
data obtained from the ESO Science Archive Facility under request number rdutta129485;
data obtained with VLBA (Prgrm. ID: BD187, PI: Dutta)}}  
\author[Dutta et al.]{R. Dutta$^{1}$\thanks{E-mail: rdutta@iucaa.ernet.in}, R. Srianand$^{1}$, S. Muzahid$^{2}$, N. Gupta$^{1}$, E. Momjian$^{3}$, J. Charlton$^{2}$ \\ 
$^{1}$ Inter-University Centre for Astronomy and Astrophysics, Post Bag 4, Ganeshkhind, Pune 411007, India \\
$^{2}$ The Pennsylvania State University, 525 Davey Lab, University Park, State College, PA 16802, USA \\
$^{3}$ National Radio Astronomy Observatory, 1003 Lopezville Road, Socorro, NM 87801, USA \\   
} 
}
\date{Accepted. Received; in original form }
\maketitle
\label{firstpage}
%
%============================== ABSTRACT =================================================================================
%
\begin {abstract}  
\par\noindent
We present a detailed analysis of a \h2-bearing metal-rich sub-damped \lymana system at \zabs = 0.10115 towards 
the radio-loud quasar J0441$-$4313, at a projected separation of $\sim$7.6 kpc from a star-forming galaxy. The \h2, $\con$ 
and $\nao$ absorption are much stronger in the redder of the two components seen in the Hubble Space Telescope / Cosmic Origins 
Spectrograph spectrum. The best single component fit to the strong \h2 component gives log~\Nh2 = 16.61 $\pm$ 0.05. However, 
possible hidden saturation in the medium resolution spectrum can allow for log~\Nh2 to be as high as 18.9. The rotational excitation 
temperature of \h2 in this component is 133$^{+33}_{-22}$ K. Photoionization models suggest 30-80\% of the total \nhi is associated 
with the strong \h2 component, that has a density $\le$ 100 \cc~and is subject to a radiation field that is $\le$ 0.5 times the Galactic 
mean field. The Very Large Baseline Array 1.4 GHz continuum image of the radio source contains only 27\% of the arcsecond scale 
emission. Using a previously published spectrum, no \21 absorption is found to be associated with the strong \h2 component. This 
suggests that either the \nhi associated with this component is $\le$ 50\% of the total \nhi or the gas covering factor is $\le$ 0.27. 
This is consistent with the results of the photoionization model that uses UV radiation due to stars in the associated galaxy. The 
\21 absorption previously reported from the weaker \h2 component suggests a spin temperature of $\le$ 90 K, at odds with the weakness 
of \h2, $\con$ and $\nao$ absorption in this component. From the inferred physical and chemical conditions, we suggest that the gas 
may be tracing a recent metal-rich outflow from the host-galaxy.
\end {abstract}  
%
%=========================== KEY WORDS =================================================================================== 
%
%
%=========================== INTRODUCTION ================================================================================== 
%
\section{Introduction} 
\label{sec_intro}  
Damped \lymana systems (DLAs) and sub-DLAs are by definition absorbers with neutral hydrogen column density, 
\nhi $\ge$ 2$\times$10$^{20}$\,cm$^{-2}$ and $\ge$ 1$\times$10$^{19}$\,cm$^{-2}$ respectively \citep[see][for a review]{wolfe2005}. 
These high $\hon$ column density absorbers trace the bulk of the $\hon$ at 2 $<z<$ 3 \citep{peroux2005,noterdaeme2009b,noterdaeme2012}, 
and can in principle contribute significantly to the global star formation rate \citep{wolfe2003,srianand2005}. 
While they have been conjectured to be originating from gas associated with high-$z$ galaxies/proto-galaxies, 
their direct connection with galaxies is not yet well understood. The link between DLAs/sub-DLAs and galaxies 
can be established by directly detecting the host galaxies and/or showing that the prevailing physical 
conditions in the absorbing gas are consistent with those seen in a typical galactic interstellar medium (ISM). 

Due to difficulties associated with detecting high-$z$ galaxies that are close to bright QSO 
sightlines, our understanding of the physical conditions in DLAs/sub-DLAs at $z$ $>$ 1.8 is 
primarily based on optical absorption line studies of low-ionization metal transitions and, 
in a few cases, \h2, HD and CO molecular transitions \citep[e.g.][]{varshalovich2001,ledoux2003,noterdaeme2008a,noterdaeme2008b,noterdaeme2009a,srianand2008}. 
The observed high-$J$ excitations of \h2 at high-$z$ are consistent with a strong ultraviolet 
(UV) field suggesting in-situ star formation, while the typical temperatures and densities of the 
\h2 components are 153 $\pm$ 78 K and 10$-$200\,cm$^{-3}$ respectively \citep{srianand2005}. 
The gas producing \h2 absorption probably traces diffuse molecular gas in the form of compact 
clouds \citep{balashev2011} and contains only a small fraction of the $\hon$ measured using the 
DLA profile \citep{srianand2012}.  

The $\hon$ \21 absorption line provides a complementary way to probe the physical conditions in 
the $\hon$ gas. If detected it can be used to investigate: (1) the thermal state of $\hon$ gas, 
as \21 optical depth is a good tracer of the gas kinetic temperature being inversely proportional 
to the spin temperature \citep{kulkarni1988}; (2) the parsec-scale structure in the absorbing gas 
via milliarcsecond (mas) scale spectroscopy \citep{srianand2013}; (3) the magnetic field in the
cold neutral medium (CNM) using Zeeman splitting \citep{heiles2004}; (4) the filling factor 
of cold gas in the ISM. Unfortunately, \21 detections from DLAs at high-$z$ are very rare 
\citep[see][for recent compilations]{srianand2012,kanekar2014}. This lack of \21 detection 
is attributed to the gas being warm and/or to partial coverage of the background radio sources 
by the absorbing clouds. 

The effect of covering factor and the size of absorbing clouds can in principle be 
quantified by searching for \21 absorption in DLAs/sub-DLAs with \h2 detections. In 
our Galactic high latitude sightlines, a correspondence between \h2 and \21 absorption 
is well established when log~\Nh2 $\ge$ 15.8 \citep[see][]{roy2006}. Based on this study, 
we know that the gas kinetic temperature measured from the rotational level populations 
of \h2, \th, traces the $\hon$ spin-temperature, \ts. Therefore, in that case the lack of 
\21 absorption from a gas showing low rotational temperature of \h2 cannot be naturally 
attributed to the gas being warm. However, high-$z$ DLAs with \h2 detection towards radio-bright 
QSOs (with flux density in excess of 100 mJy at the redshifted \21 frequencies), at appropriate 
\zabs (enabling \21 absorption search) are rare. Even the five cases where this is possible, 
do not follow the relationship between \21 and \h2 absorption \citep[see][]{srianand2012}. 
This lack of correspondence may mean that only a small fraction (i.e., $\le$ 10\%) of the 
total $\hon$ is associated with the \h2 components or that the \h2 clouds are too small 
(i.e., $<$ 15\,pc) to cover even the unresolved radio components at mas scales. To resolve 
this, mas scale very long baseline interferometry (VLBI) imaging of the background radio 
source is essential to determine the covering factor \citep[see for example][]{gupta2012}. 
A good understanding of these issues is important to interpret the results of on-going 
and future \21 absorption line surveys, as well as surveys of \h2 in DLAs/sub-DLAs.

Low-$z$ DLAs and sub-DLAs are ideal targets to address the above issues. This is because:
(1) for a given angular scale one will be probing a much smaller physical scale, where 
the covering factor issues may have minimum effects \citep[see][]{curran2005}; (2) in 
case of \21 detections one will be able to carry out VLBI spectroscopic observations 
to get spatial scales in the absorbing $\hon$ gas \citep{borthakur2010,srianand2013}; 
(3) direct association of the absorbers with the galaxies responsible for the absorption 
is possible. The last one will allow us to relate the inferred spatial scales in the 
absorbing gas to its location with respect to the galaxy (i.e., in the star forming regions 
or in the halo). \citet{srianand2013} have demonstrated the presence of structures in 
the \21 optical depth on parsec scales in the ISM of a low-$z$ galaxy. In such cases of 
the background radio source being structured, the spatial variation of the $\hon$ gas 
opacity can be studied using VLBI spectroscopy. 

UV spectroscopic surveys with the Hubble Space Telescope (HST), are identifying 
significant numbers of low-$z$ DLAs and sub-DLAs \citep{rao2011,meiring2011,battisti2012}. 
Thanks to the Cosmic Origins Spectrograph (COS) on board the HST, there have been three 
recent cases of \h2 detection in low-$z$ (i.e., $z$ $<$ 0.6) DLAs \citep{crighton2013,oliveira2014,srianand2014}. 
Recently by a careful search in 27 low-$z$ (i.e., $z$ $<$ 1) DLAs and sub-DLAs towards 
26 QSOs, using HST/COS archival spectra, \citet{muzahid2014} have reported 7 new \h2 
detections. With the largest sample to date (10) of \h2 detections at low-$z$, they find 
$\gtrsim$ 2 times higher \h2 incidence rate at low-$z$ compared to high-$z$. The \h2 
components in their sample have typical temperatures of 133 $\pm$ 55 K. From photoionization 
models and the lack of high-$J$ excitations, the authors have shown that the prevailing 
radiation field is much weaker than the Galactic UV radiation field, in contrast with the 
results at high-$z$. Moreover, the large impact parameters of the host-galaxies ($>$ 15 kpc) 
for majority of their systems strongly suggest that the \h2 bearing gas is not related to 
star-forming disks but probably stems from extended halo gas.

Among the low-$z$ sample of \h2 absorbers of \citet{muzahid2014}, two are towards radio-loud 
QSOs, which in principle facilitates detailed study of the cold gas in these absorbers. However, 
the redshift of one of these systems falls in the frequency range affected by radio frequency 
interference at available radio telescopes. The other system is the subject of detailed 
study in this paper. This paper is organized as follows. In Section~\ref{sec_obs}, we give details 
of the observations and data reduction process. The physical properties of the system as derived 
from the absorption lines are discussed in Section~\ref{sec_analysis}. The properties of the 
associated galaxy and their comparison with that of the absorber are discussed in Section~\ref{sec_galaxy}. 
Section \ref{sec_cloudy} explains the results from our photoionization modelling of the absorber. 
In Section~\ref{sec_radio}, we focus on the Very Long Baseline Array (VLBA) mas scale imaging of 
the background radio source at 1.4 GHz and the previously reported \21 absorption from the cold gas 
using the Australia Telescope Compact Array (ATCA) . Lastly, we summarize our results in Section~\ref{sec_sum}. 
Throughout this paper we use a flat cosmology with H$_{0}$ = 70 \kms Mpc$^{-1}$ and $\Omega_{m}$ = 0.27.
%  
%==================================================== OBSERVATIONS =========================================================
%
\section{Observations and data reduction}  
\label{sec_obs}  
The QSO J044117.3$-$431343 (\zem = 0.59378; also known as PKS0439$-$433 and henceforth J0441$-$4313) 
was observed using COS during HST cycle-19, under program ID: 12536 (PI: V. Kulkarni). These observations 
consist of G130M far-UV grating integrations (5.3 ks) at a medium resolution of R $\sim$ 18000 
(full width at half maximum, FWHM $\sim$ 18 \kms). The data were retrieved from the HST 
archive and reduced using the CALCOS pipeline software. The individual reduced $x1d$ files 
were first flux calibrated. Then the alignment and co-addition of the separate exposures 
were carried out using the software developed by \citet{danforth2010}\footnote{http://casa.colorado.edu/∼danforth/science/cos/costools.html}. 
The exposures were weighted by the integration time while co-adding the flux calibrated 
data. The final co-added spectrum covers the observed wavelength range 1131$-$1435 \AA~ and 
has a signal-to-noise ratio, S/N $\sim$ 16$-$20 per resolution element. Since each COS 
resolution element is sampled by six raw pixels, we binned the spectrum by 3 pixels, 
which further improves the S/N per pixel. All measurements and analyses were, subsequently, 
performed on the binned data. Measurements are, however, found to be fairly independent 
of the binning. Continuum normalization was done by fitting the line free regions with a 
smooth lower order polynomial. 

The COS wavelength calibration is known to be uncertain at the level of 10$-$15 \kms
\citep{savage2011,meiring2013}. Regions of the spectrum that are recorded near the edges 
of the detector segment are likely to have an erroneous wavelength solution
\citep[see for e.g.][]{muzahid2014}. However, in the case of the present spectrum, the velocity 
offsets at different wavelengths as estimated from the difference in the observed and the 
expected line centroids of the numerous \h2 absorption lines are $\lesssim$ $\pm$5 \kms. 
Moreover, the Galactic absorption lines are found to be aligned within $\lesssim$ $\pm$5 \kms. 
The line spread function (LSF) of the COS spectrograph is not a Gaussian. \citet{ghavamian2009} 
provided a characterization of the non-Gaussian LSF for COS, which was subsequently updated 
by \citet{kriss2011}. We adopt this updated LSF for our Voigt profile fitting analysis. 
Interpolated LSF at the line centre was convolved with the model Voigt profile while fitting 
an absorption line using the \textsc{vpfit}\footnote{http://www.ast.cam.ac.uk/$\sim$rfc/vpfit.html} code.

In addition, we use the pipeline calibrated spectrum (R $\sim$ 45000, FWHM $\sim$ 6.6 \kms) 
of QSO \sys obtained using the Ultraviolet Echelle Spectrograph (UVES) on the Very Large 
Telescope (VLT), available in the European Southern Observatory (ESO) archive 
\footnote{http://archive.eso.org/wdb/wdb/adp/phase3\_spectral/form}. 
The spectra corresponding to different exposures covering our regions of interest, after applying 
barycentric correction, were brought to their vacuum values using the formula given in \citet{edlen1966}. 
For the co-addition, we interpolated the individual spectra and their errors to a common wavelength 
array, and then computed the weighted mean using the weights estimated from the error in each pixel. 
The final spectrum covers the observed wavelength range 3732$-$6837 \AA~ with S/N $\sim$ 40$-$50 
per resolution element.

Furthermore, we observed the background radio source with the VLBA using the \21 receiver band 
for 1 hr on 2014 August 26. The total bandwidth was 256 MHz in dual polarization (eight 32 MHz 
baseband channel pairs). Each baseband channel was split into 128 spectral points. Two-bit 
sampling and a correlator integration time of 2 s were used. The observations were carried out 
in nodding-style phase referencing with a cycle of $\sim$5 min, i.e., $\sim$3.5 min on source 
and $\sim$1.5 min on the phase calibrator (J0440$-$4333). A strong fringe finder/bandpass 
calibrator (DA193) was also observed at the beginning for 4$-$5 min. The target source was 
observed for $\sim$35 min, split into scans at different hour angles to improve the UV-coverage. 
Data were calibrated and imaged using Astronomical Image Processing System \citep[\textsc{aips};][]{greisen2003} 
in a standard way \citep[see for example][]{momjian2002,srianand2012}.
%
%=================================================== ANALYSIS ============================================================== 
%
\section{Absorption Line Properties of the sub-DLA}
\label{sec_analysis}
The sub-DLA at \zabs = 0.10115 towards \sys~with log~\nhi = 19.63 $\pm$ 0.08 and log~\Nh2 = 16.61 $\pm$ 0.05 
\citep{muzahid2014} presents a very interesting case for understanding the cold gas that is present around galaxies. 
This system was originally selected as a DLA candidate based on strong $\mgt$ and $\fet$ absorption in the 
Faint Object Spectrograph (FOS) spectrum \citep{petitjean1996}. We note that \citet{chen2005} derived log~\nhi = 19.85 $\pm$ 0.10 
for this sub-DLA using the Space Telescope Imaging Spectrograph (STIS) G140L spectrum, consistent within 2$\sigma$ 
of our measurement. In this section, we discuss the properties of the sub-DLA as probed by the absorption lines 
(metals and \h2) detected in the COS and UVES spectra. 
%
%=========================== Discussion on Metallicity =====================================================================
%
\subsection{Analysis of metal lines} 
\label{sec_metal}
\begin{figure*}
\includegraphics[width=0.4\textwidth, angle=90]{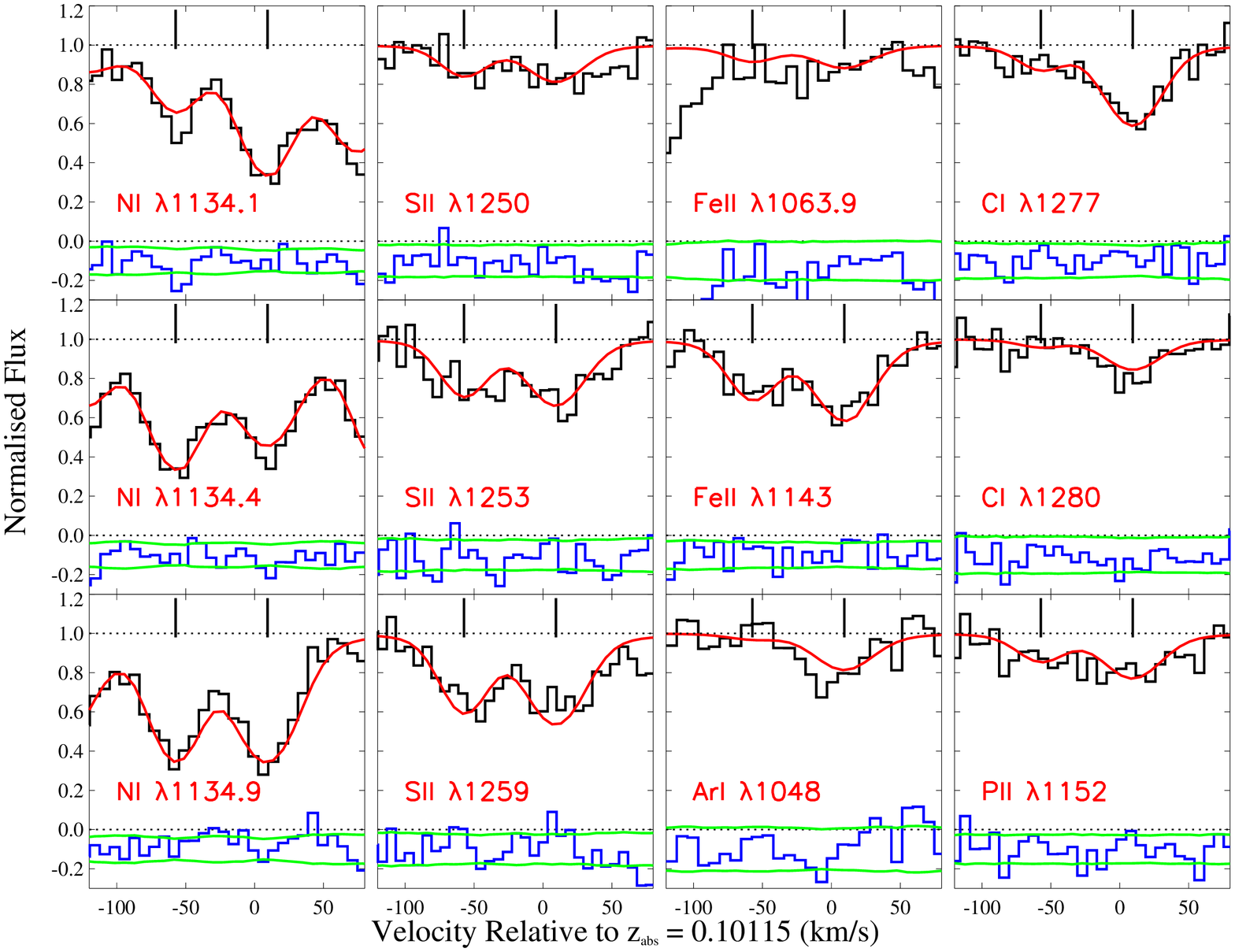}
\includegraphics[width=0.3\textwidth, angle=90]{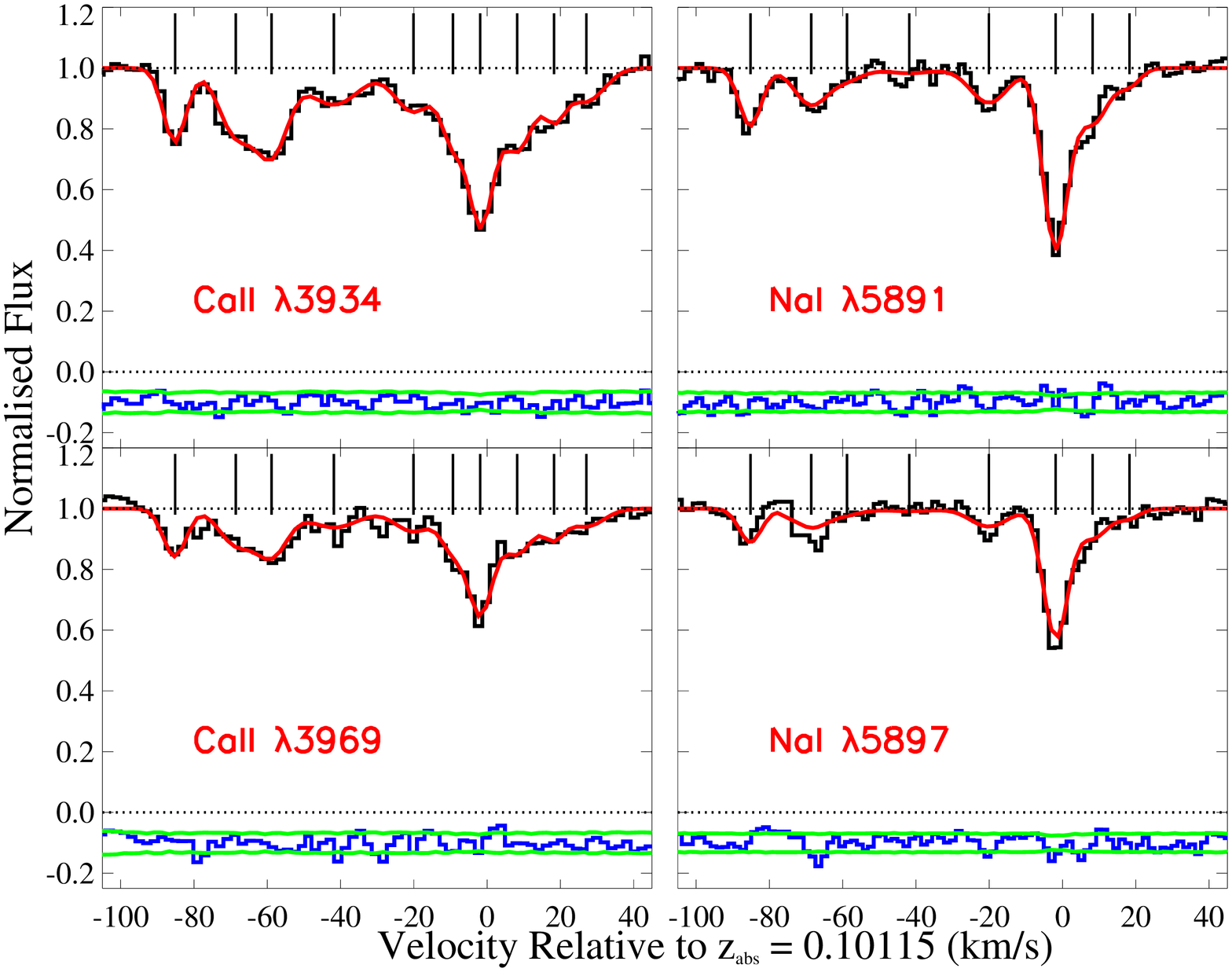}
\caption{A selection of metal lines associated with the sub-DLA at \zabs = 0.10115 towards \sys detected in the 
COS (\textit{Left}) and UVES (\textit{Right}) spectrum. Best-fitting Voigt profiles are overplotted in red. The 
tick marks show the component positions. The errors in flux and residuals from the fit are shown at the bottom 
as green lines and blue histograms respectively.}
\label{fig_vplot}
\end{figure*}

We detect the metal transitions $\con$, $\cto$, $\non$, $\nto$, $\oon$, $\sit$, $\sitr$, $\pto$, 
$\sto$, $\aro$, $\fet$ and $\fetr$ in the COS spectrum spread over a velocity range of $\sim$150 \kms. 
Some of these absorption profiles are shown in the left panel of Fig.~\ref{fig_vplot}. The $\cto$, 
$\nto$, $\oon$, $\sit$, $\sitr$ and $\fetr$ lines are saturated/heavily blended and hence not used 
in our analysis. We do not cover the $\cto$* \ll1335.7, 1335.6 transitions, and the $\cto$* \l1037 
transition is heavily blended. $\con$* absorption is absent and we place a 3$\sigma$ upper limit on 
its column density using the strongest transition that is free from any blend, i.e., \l1279.9. Further,
$\mgt$ \ll1239, 1240 and $\mnt$ \l1197 absorption are not detected and we use these to place 3$\sigma$ 
upper limits on $N$($\mgt$) and $N$($\mnt$) respectively. From the Voigt profile fits to the unblended 
and most likely unsaturated metal lines detected in the COS spectrum, we infer two main absorbing clumps 
of gas at \zabs = 0.10094 \& 0.10119 (henceforth Component 1 and 2 respectively), $\sim$ 68 \kms~apart 
(see left panel of Fig.~\ref{fig_vplot}). We assume that all the neutral and singly ionized atoms are physically 
associated with the same gas cloud. Hence while fitting, the redshift and $b$ parameter for each absorption 
component are tied to be the same for each of the ions, i.e., only turbulent broadening is considered. 
Additionally, we rejected fits with more than two components since, although having similar \chin, they 
have larger values of AICC (Akaike information criteria \citep[][]{akaike1974} corrected for the finite 
sample size \citep[AICC;][]{sugiura1978} as given in equation 4 of \citet{king2011}) and higher errors
in the parameters.  

The medium resolution of the COS spectrum is not sufficient to resolve all the different components in 
the absorbing gas, in particular the narrow ones. In the present case, since $\cat$ and $\nao$ absorption 
lines have been detected in the high resolution UVES spectrum \citep{richter2011}, we can get a 
realistic idea of the component structure. We find that ten and eight components give the best fit to 
the $\cat$ and $\nao$ lines respectively (see Table~\ref{tab_uves} and right panel of Fig.~\ref{fig_vplot}). 
The column densities that we obtained are consistent within errors with those reported by \citet{richter2011}. 
Table~\ref{tab_colden} lists the column density measurements or 3$\sigma$ upper limits in case of 
non-detections of the various ions. In the case of $\cat$ and $\nao$, to facilitate comparison between 
Components 1 and 2 seen in the COS spectrum, we sum the column densities in components (a) 
to (d) and (e) to (j) respectively. 

Both the $\aro$ lines (\l1066 and \l1048) are not detected in Component 1. Absorption 
from Component 2 at the expected wavelength ranges of $\aro$ is present. The $\aro$ 
\l1066 line is blended with some other absorption, and the $\aro$ \l1048 line centroid is shifted 
by 1 pixel with respect to the expected peak absorption from other metal lines (see Fig.~\ref{fig_vplot}). 
It is not clear whether this shift is caused by wavelength scale uncertainties or not. Regardless, 
it is clear that the $\aro$ strength we find is lower than what we expect if there is no relative 
depletion between Ar and S and if ionization corrections are ignored. Later we will discuss 
this issue in detail.
 
Since \nhi cannot be decomposed into components, the average abundance can be estimated as, 
[X/H] = log~($N$(X \textsc{n})/$N$($\hon$)) $-$ log~($N$(X)/$N$($\hon$))$_{\odot}$ + IC, where 
$N$(X \textsc{n}) is the sum of the column densities of all the components of the dominant ion 
\textsc{n} of element X, and IC is the average ionization correction. Throughout this paper we 
use the solar elemental abundances as given in \citet{asplund2009}. In Table~\ref{tab_colden}, 
we list the average metallicities without any ionization corrections for those species which 
are expected to be the dominant ions of the respective elements in the neutral phase. IC for 
the S abundance, determined from $\sto$, is found to be negligible compared to the errors as 
discussed in Section~\ref{sec_cloudy}. Here, IC = log~$\rm f_{\rm H\,{\sc I}} / f_{\rm S\,{\sc II}}$, 
where $\rm f_{\rm H\,{\sc I}}$ and $\rm f_{\rm S\,{\sc II}}$ are the ionization fractions of 
$\hon$ and $\sto$ respectively. Moreover, since S is known not to be affected by dust depletion, 
we can take the average metallicity of the absorber as that of the S abundance, which is found 
to be super-solar ([S/H] = 0.28 $\pm$ 0.08). Additionally, the abundance of P, most likely a 
non-refractory element, is found to be super-solar. This is higher than the previously 
reported value of log~$Z$ = $-$0.20 $\pm$ 0.30 by \citet{chen2005}, based on the $\fet$ column 
density measured using the low-resolution FOS spectrum, log~\nhi = 19.85 $\pm$ 0.10, and the 
mean measured value of $N$($\znt$)/$N$($\fet$) in the DLA population. 

When we consider the total column densities, we find [Ar/S] = $-$1.04 $\pm$ 0.14. The ratio [Ar/S] is 
$\le$ $-$1.66 and $-$0.85 in Components 1 and 2 respectively. This is much lower than the mean value of 
$-$0.4 found by \citet{zafar2014} for high-$z$ DLAs. The large Ar depletion we find towards our high 
metallicity system with low \nhi is in line with the mild correlations of the [Ar/$\alpha$] ratio
with metallicity and \nhi as noted by \citet{zafar2014}. We wish to point out that such large $\aro$ 
depletions are seen in the case of High Velocity Clouds (HVCs) in the Milky Way \citep[see][]{richter2001}. 
We discuss this issue in detail in Section~\ref{sec_cloudy}. 

$\fet$ is the only iron co-production species clearly detected in our spectrum. In the absence of a $\znt$ 
column density measurement it is difficult to interpret the Fe abundance, since differences between Fe and 
any other $\alpha$ element may reflect either Fe depletion or nucleosynthetic origin. Here we proceed with 
the assumption that the lower abundance of Fe is a reflection of dust depletion. Hence, the dust depletion 
obtained is, [Fe/S] = $-$0.49 $\pm$ 0.04, the column density of dust in Fe is, \ln(Fe$_{\rm dust}$) = 15.24 $\pm$ 0.01, 
and the dust-to-gas ratio is, $\kappa$ = 1.29 $\pm$ 0.29\footnote{Here, $N(\rm Fe_{\rm dust}) = N(\rm S)[ 1 - 10^{[\rm Fe/S]}](\rm Fe/S)_{\odot}$ 
\newline and, $\kappa = 10^{[\rm S/H]} [ 1 - 10^{[\rm Fe/S]}]$}. The inferred Fe depletion is much less than 
what is typically seen in the cold gas in the Galactic ISM. From Fig.~9 of \citet{welty1996}, we notice that 
in the Galactic ISM, $N$($\nao$)/$N$($\cat$) $>$ 1 for the observed values of $N$($\nao$) in the present system. 
On the contrary, we find that $N$($\cat$) is higher than $N$($\nao$) in the present case. This could mean Ca 
depletion is not as high as typically seen ($\sim$ $-$3 dex) in the Galactic ISM. Such a picture is consistent 
with relatively small depletion we infer for Fe from the observed [Fe/S]. As noted previously, we do not detect 
$\mgt$ \ll1239, 1240 and $\mnt$ \l1197. However, the 3$\sigma$ limits we obtain are not stringent enough to 
provide further insights into the chemical history of this system.

High-$z$ DLAs with such high values of metallicities, $N(\rm Fe_{\rm dust})$ and $\kappa$ tend to show 
\h2 absorption \citep{ledoux2003,petitjean2006,noterdaeme2008a}. \h2 molecular absorption is indeed detected 
from this system at \zabs = 0.10115 from $J$ = 0, 1, 2, 3 levels by \citet{muzahid2014}. We also notice 
a consistent albeit weaker \h2 absorption at \zabs = 0.10091. From equivalent width measurements of 
$\mgt$ \ll2796, 2803, $\fet$ \l2600 given in \citet{churchill2001} using the FOS spectrum, and the results 
of \citet{gupta2012}, we find that there is a high probability of detecting \21 absorption from this system. 
A tentative weak \21 absorption has been reported from this system at \zabs = 0.10097, however no \21 
absorption corresponding to the stronger \h2 component (i.e., Component 2) is detected \citep{kanekar2001}. 
In Section~\ref{sec_21cm}, we will discuss the implications of this \21 measurement in detail.

We do not find any strong (i.e., $>$ 3$\sigma$) difference in the ratios $N$($\fet$)/$N$($\sto$), 
$N$($\pto$)/$N$($\sto$) and $N$($\non$)/$N$($\sto$) between the two components. While most of the metal column 
densities in the two components match within $\sim$0.2 dex, $\con$, $\nao$, $\aro$ and \h2 absorption are much 
weaker in Component 1. Since these species (apart from $\aro$) can be ionized by photons with energy $\lesssim$ 
11 eV, their column densities will be useful in constraining the ionizing radiation field, as explored in Section 
\ref{sec_cloudy}. 

From the column densities of $\non$, $\pto$, $\sto$, $\cat$, $\fet$, we find that $\sim$ 55$-$65\% of the 
total metal column density is in Component 2. This implies that the fraction of \nhi in the strong \h2 
component is, $f_{\rm N(H\,\textsc{i})}$ $\sim$ 0.55$-$0.65, if the metallicity is uniform across the components. 
On the other hand, if $N$($\nao$) scales with \nhi as seen in our Galaxy \citep{ferlet1985,wakker2000}, 
then we would expect $\sim$80\% of the total \nhi to be present in Component 2. Hence, from the observed 
distribution of the metal column densities across the two components, $f_{\rm N(H\,\textsc{i})}$ is expected 
to be $\sim$ 0.55$-$0.8 in Component 2. If the known relations of $N$($\nao$) and $N$($\cat$) with \nhi in our 
Galaxy \citep{ferlet1985,wakker2000} were valid for the present system, we would expect \ln($\nao$) = 11.49 
and \ln($\cat$) = 11.77. However, the observed column densities are ten times higher. In other words, 
$\sim$10 times more $\nao$ and $\cat$ per $\hon$ are present in this sub-DLA than what is typically seen 
in the Galactic ISM. As both $\nao$ and $\cat$ are not the dominant ions of the respective elements in 
the $\hon$ phase, this may imply that the background ionizing field in this sub-DLA is weaker than the 
mean Galactic radiation field. We come back to this point in Section~\ref{sec_cloudy}.
\begin{table} 
\caption{Component-wise column densities (with errors shown in parentheses) of $\cat$ and $\nao$ detected from the sub-DLA at \zabs = 0.10115 towards \sys}
\centering
\begin{tabular}{ccccc}
\hline
\hline
Comp. & \zabs              & $b$         & \ln($\cat$) & \ln($\nao$)   \\
      &                    & (\kms)      & (\cms)      & (\cms)        \\
\hline
a     & 0.10084 (0.000001) & 1.55 (0.46) & 11.61 (0.04) & 11.26 (0.03) \\
b     & 0.10090 (0.000003) & 4.91 (1.03) & 11.63 (0.09) & 11.16 (0.06) \\
c     & 0.10093 (0.000003) & 4.44 (1.08) & 11.74 (0.08) & 10.59 (0.25) \\
d     & 0.10100 (0.000005) & 8.80 (2.34) & 11.53 (0.08) & 10.46 (0.23) \\
e     & 0.10108 (0.000002) & 5.88 (0.93) & 11.48 (0.06) & 11.19 (0.04) \\
f     & 0.10112 (0.000002) & 1.55 (0.87) & 11.54 (0.06) & ---          \\
g     & 0.10114 (0.000001) & 2.19 (0.20) & 12.08 (0.03) & 12.04 (0.02) \\
h     & 0.10118 (0.000002) & 4.37 (1.16) & 11.72 (0.05) & 11.33 (0.06) \\
i     & 0.10122 (0.000005) & 0.84 (0.80) & 11.37 (0.16) & 10.67 (0.16) \\
j     & 0.10125 (0.000007) & 5.35 (2.56) & 11.34 (0.13) & ---          \\
\hline
\hline
\end{tabular}
\label{tab_uves}
\end{table}
\begin{table*} 
\caption{Component-wise column densities (with errors shown in parentheses) of metals detected from the sub-DLA at \zabs = 0.10115 towards \sys}
\centering
\begin{tabular}{ccccc}
\hline
\hline
Ion (X \textsc{n}) & \multicolumn{3}{c}{\ln (\cms)}     & [X \textsc{n}/H]$^{c}$              \\ 
\cline{2-4}   & Component 1$^{a}$  & Component 2$^{b}$  & Total              &                \\
\hline
$\hon$        & ---                & ---                & 19.63 (0.08)       & ---            \\
$\con$        & 13.31 (0.14)       & 14.01 (0.04)       & 14.09 (0.04)       & ---            \\     
$\con$*       & $\le$ 13.40        & $\le$ 13.50        & $\le$ 13.75        & ---            \\
$\non$        & 14.64 (0.02)       & 14.71 (0.02)       & 14.98 (0.02)       & $-$0.48 (0.08) \\  
$\nao$ $^{1}$ & 11.60 (0.04)       & 12.18 (0.02)       & 12.28 (0.02)       & ---            \\ 
$\mgt$        & $\le$ 15.30        & $\le$ 15.30        & $\le$ 15.60        & $\le$ 0.37     \\    
$\pto$        & 13.00 (0.11)       & 13.29 (0.07)       & 13.47 (0.06)       & $+$0.43 (0.10) \\
$\sto$        & 14.65 (0.04)       & 14.79 (0.03)       & 15.03 (0.02)       & $+$0.28 (0.08) \\                 
$\aro$        & $\le$ 12.27        & 13.22 (0.12)       & 13.27 (0.14)       & $-$0.76 (0.16) \\  
$\cat$ $^{1}$ & 12.24 (0.04)       & 12.45 (0.03)       & 12.66 (0.02)       & ---            \\
$\mnt$        & $\le$ 12.80        & $\le$ 12.90        & $\le$ 13.2         & $\le$ 0.14     \\
$\fet$        & 14.49 (0.05)       & 14.72 (0.03)       & 14.92 (0.03)       & $-$0.21 (0.09) \\
\hline
\hline
\end{tabular}
\label{tab_colden}
\begin{flushleft}
$^{a}$ \zabs = 0.10094 $\pm$ 0.000003 \& $b$ = 19.6 $\pm$ 1.5 \kms \\
$^{b}$ \zabs = 0.10119 $\pm$ 0.000003 \& $b$ = 23.8 $\pm$ 1.3 \kms \\ 
$^{c}$ Average metallicities without applying any ionization corrections \\
$^{1}$ The column densities in Components 1 \& 2 are summed over components (a) to (d) \& (e) to (j) 
       respectively as listed in Table \ref{tab_uves} 
\end{flushleft}
\end{table*}
%
%=========================== Discussion on H2, CI and CI* ==================================================================
%
\subsection{Physical Conditions} 
\label{sec_physical}
\subsubsection{\h2 absorption}
\label{sec_h2}
Details of the \h2 absorption properties in the two components are given in Table~\ref{tab_h2}, while 
the fits are shown in Fig. C2 of \citet{muzahid2014}. We consider the \Nh2 obtained from \textsc{vpfit} 
for the weaker component as an upper limit as most of the transitions are blended. Both the components 
are $\sim$10 \kms~shifted from the redshifts of the corresponding metal components derived using the COS 
spectrum. However, the stronger \h2 absorption arises within $\sim$2 \kms~of the strongest components 
of $\cat$ and $\nao$. The inferred excitation temperature is $\le$ 200 K in both the components, similar 
to what is expected from a CNM gas. However, this gas is situated outside the visible optical disk of 
the candidate host-galaxy \citep[see Fig.~1 of][]{chen2005}. If we consider the \Nh2 of the weaker 
component as a measurement, the total \h2 column density for this system is log~\Nh2 = 16.64 $\pm$ 0.05, 
and hence the average molecular fraction is log~f(H$_{2}$) = $-$2.69 $\pm$ 0.09\footnote{f(H$_{2}$) = 2\Nh2 $/$ (2\Nh2 $+$ \nhi)}. 
HD absorption is undetected and from the strongest unblended transition of HD($J$=0)~(\l1042), we 
estimate a 3$\sigma$ upper limit of \ln(HD,$J$=0) $\le$ 13.90.

As pointed out by \citet{muzahid2014}, due to the moderate spectral resolution of the COS spectrum, 
\Nh2 measurements can be highly uncertain if the intrinsic $b$ values are small. To test the uncertainties in
the \Nh2 measurement, we fitted the \h2 in Component 2 with multiple components with their $z$ and 
$b$ fixed to that of the $\nao$ components arising within 30 \kms~of \zabs = 0.10115. We find that 
the \h2 lines can be adequately fitted with at most three of these components, and further components are 
superfluous. Of these three components, the component corresponding to the strongest $\nao$ with 
$b$ = 2 \kms, contributes $\ge$ 99\% to the total \Nh2. The resultant log~\Nh2 = 18.9 $\pm$ 0.1 from 
this fit is much higher than that obtained from our single component fit with $b$ as a free parameter. 
While the \th~(127 $\pm$ 10 K) obtained from this fit is consistent with that obtained from our single 
component fit. However, the multi-component fit gives higher AICC value (1675) as well as larger parameter 
errors compared to our single component fit.

Note that our approximation of fixing the $b$ of the \h2 lines as that of the $\nao$ components 
is not realistic, especially if the thermal broadening is large. In addition, as the strongest $\nao$ 
component contains almost all the \Nh2, we can approximate the \h2 absorber as a single cloud, and use 
the curve of growth (COG) to get a better handle on the $b$ parameter of \h2. For details of the COG 
technique to estimate the $b$ of \h2 transitions see \citet{srianand2014}. We notice that the absorption 
lines of the $J$ = 3 transitions from Component 2 which have a spread in the $\lambda f$ values ($\lambda$: 
rest wavelength, $f$: oscillator strength), show a wide range of rest-equivalent widths. Hence using a single 
cloud COG for \h2($J$=3) transitions, we obtain $b$ = 4$-$7 \kms~for Component 2. This is of the order of 
one-third to one-half of the resolution of the COS spectrum. We then fit all the \h2 lines by fixing $b$ to 
be 4$-$7 \kms~in \textsc{vpfit}. Here we have assumed that the $b$ values of different $J$ levels are the 
same \citep[see however][]{noterdaeme2007}. These fits give log~\Nh2 = 18.70$-$18.90 for Component 2, consistent 
with the multi-component fit with $b$ fixed to that of the $\nao$ components, and \th~= 113 $\pm$ 10 K. However, 
the AICC for these fits are larger compared to the fit in which the $b$ is a free parameter, indicating that the 
present data do not support a narrow $b$ and a high \Nh2. Moreover, \h2 absorption from the higher $J$-levels 
($J\ge$ 4) as well as HD absorption, as expected for such large \Nh2, are not detected in the present spectrum. 
However, we note that a higher resolution spectrum is required to accurately estimate both these quantities. 
Here we proceed assuming log~\Nh2 = 16.61 $\pm$ 0.05 for \h2 in Component 2, though we explore the possibility 
of the gas having high \Nh2 using photoionization modelling in Section~\ref{sec_highnh2}. The important point 
to note here is that the \th~does not seem to depend on the $b$ parameter, and both the multi-component fit 
and the single cloud COG approach lead to \th~consistent with our original estimate.

We estimate a 3$\sigma$ upper limit of log~$N$(\h2,$J$=4) $\le$ 13.82, using the strongest unblended line 
(\l1044) available. This allows us to constrain the photoabsorption rate of Lyman- and Werner-band 
UV photons by \h2 in the $J$ = 0 level ($\beta_{0}$) from the equilibrium of the $J$ = 4 level population. 
From equation (5) and values given in \citet[][Section~4.2]{noterdaeme2007}, we get $\beta_{0}$ $\le$ 2 
$\times$ 10$^{-11}$ s$^{-1}$, i.e., one-tenth of the Galactic rate or lower. From simple formation 
equilibrium of optically thin \h2 \citep{jura1975}, we have the neutral hydrogen density, \nh~= 
0.11$\beta_{0}$\Nh2$/(R$\nhi$)$, where $R$ is the formation rate of \h2. We use the ISM value of $R$ 
(3 $\times$ 10$^{-17}$ s$^{-1}$ \cc) scaled by the measured dust content in the system, to obtain 
the average density of the gas, \nh~$\le$ 50 \cc. Hence, from the \h2 formation, the average gas 
density is constrained to be less than $\sim$50 \cc, if the values of relevant parameters are similar 
to those measured in the Milky Way.
\subsubsection{Ionization and fine-structure excitation of C$^{0}$}
\label{sec_carbon}

Assuming solar relative abundance of C and S, and that $\cto$ in the neutral phase can be traced 
by $\sto$, we can obtain $N$($\cto$). Under this assumption, we derive some of the physical conditions 
in Component 2 using standard techniques. From photoionization equilibrium between $\con$ and $\cto$ 
and taking the gas temperature as \th, we can estimate the electron density, \ne~\citep[][Equation (5)]{srianand2005}. 
Assuming the Galactic photoionization rate for $\con$ ($\Gamma$ = 2$-$3.3 $\times$ 10$^{-10}$ s$^{-1}$; \citet{pequignot1986}) 
results in \ne~$\sim$ 0.03$-$0.10 \cc~for Component 2. If C is depleted with respect to S, we expect 
the electron density to be higher than this value. On the other hand, if the radiation field strength is 
1/10$^{th}$ of the Galactic mean field, as suggested from the lack of \h2 absorption from $J$ = 4 level, 
the inferred electron density will be lower. Taking \ne/\nh~$\approx$ 10$^{-3}$, typical of CNM gas 
\citep{wolfire1995}, leads to \nh~$\sim$ 30$-$100 \cc~in this gas for the above assumed $\Gamma$. 
Considering Component 1, the \ne/$\Gamma$ ratio is 3.6 times lower than that of Component 2. If we 
assume $\Gamma$ to be similar for both the components, this will imply that the gas density in 
Component 1 is lower than that of Component 2.

\begin{figure}
\centering \includegraphics[width=0.4\textwidth]{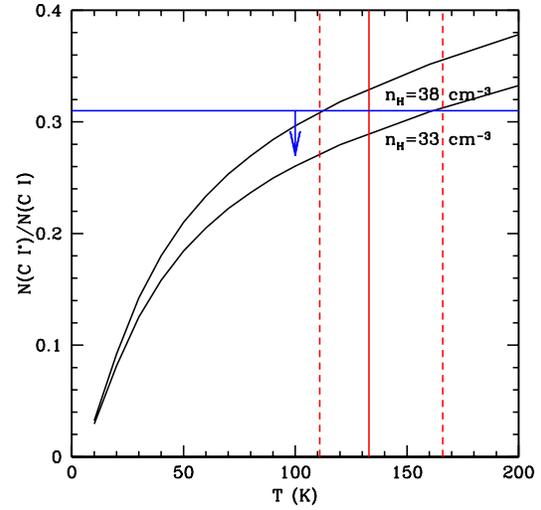} \vspace{2cm}
\caption{The ratio $N$($\con$*)/$N$($\con$) plotted as a function of temperature for different \nh. The horizontal 
line shows the observed upper limit on $N$($\con$*)/$N$($\con$), while the solid and dashed vertical lines are the 
inferred temperature and 1-$\sigma$ range respectively for the strong \h2 component.}
\label{fig_c1}
\end{figure}

In addition, from $\con$ and $\con$* equilibrium we estimate \nh~corresponding to Component 2, following 
the procedure and values given in \citet[][Section~4.2]{srianand2005}. In Fig.~\ref{fig_c1}, we plot the 
$N$($\con$*)/$N$($\con$) ratio as a function of temperature for different \nh. From the observed upper limit 
of this ratio and \th~estimated for Component 2, we get a limit on the density, \nh~$\le$ 38 \cc. While we 
have considered the Galactic UV pumping rate ($\Gamma_{01}$ = 7.55 $\times$ 10$^{-10}$ s$^{-1}$) here, we 
note that scaling the rate does not affect the result significantly, which is expected only when collisional 
excitations dominate. Therefore, the ionization and fine-structure excitation of $\con$ are consistent with 
the gas density in Component 2 being few tens of H atoms \cc. Note that the $N$($\con$*)/$N$($\con$) ratio 
measured for Component 1 is not stringent enough to place any useful constraints on the physical conditions in this gas. 
\begin{table} 
\caption{Component-wise summary of \h2 absorption properties detected in the sub-DLA at \zabs = 0.10115 towards \sys from \citet{muzahid2014}}
\centering
\begin{tabular}{ccc}
\hline
\hline
                & Component 1               & Component 2        \\
\hline
\zabs           & 0.10091 (0.000006)        & 0.10115 (0.000001) \\         
log~\Nh2 (\cms) & $\le$ 15.51 (0.03) $^{a}$ & 16.61 (0.05)       \\   
$b$ (\kms)      & 32.7 (2.6)                & 12.0 (0.5)         \\      
\th (K)         & 178 $^{+37}_{-26}$ $^{b}$ & 133 $^{+33}_{-22}$ \\      
\hline
\hline
\end{tabular}
\label{tab_h2}
\begin{flushleft}
$^{a}$ Should be treated as an upper limit as this component is severely blended \\
$^{b}$ Should be taken as indicative of the typical temperature expected in this component \\
\end{flushleft}
\end{table}
%
%=========================== Discussion on galaxy connection ===============================================================
%
\section{Connecting galaxy and absorption properties} 
\label{sec_galaxy}
The sub-DLA towards \sys is associated with a spiral galaxy at an impact parameter of 
$\sim$ 7.6 kpc from the QSO sightline \citep{petitjean1996,chen2005}. This is the lowest 
galaxy impact parameter at which \h2 absorption has been detected at low-$z$ \citep[see Fig.~9 of][]{muzahid2014}, 
enabling the connection between absorber and galaxy properties to be well-established. 
\citet{chen2005} have carried out spectroscopic analysis of the galaxy, and found it to 
be a star-forming galaxy with an oxygen abundance more than solar ([O/H] = 0.45 $\pm$ 0.15). 
They derived the sub-DLA metallicity as log~$Z$ = $-$0.20 $\pm$ 0.30 (see Section~\ref{sec_metal}), 
and hence reported a metallicity decrement along the galaxy disk.
Using higher-resolution data, however, we have found the average metallicity in the sub-DLA 
to be twice solar. Therefore, there does not appear to be any considerable metallicity 
gradient ($-$0.02 $\pm$ 0.17 dex kpc$^{-1}$) along the disk of the galaxy. This is 
similar to the results of \citet{peroux2012}, who do not find any significant decrease 
between metallicity of galaxy from emission and metallicity of absorber from absorption, 
and even report the possibility of an increasing metallicity gradient in few cases. 

Using the luminosities of the galaxy's emission lines provided by \citet{chen2005}, and 
the abundance determination method given in \citet{izotov2006}, we estimate [N/H] = $-$0.16 
and hence [N/O] = $-$0.61 in the galaxy. Here we have assumed a typical electron temperature 
of 10$^{4}$ K. The average [N/S] ratio for the sub-DLA is $-$0.76 $\pm$ 0.03 (ionization 
corrections are negligible compared to the errors). Note that this ratio is similar within 0.1 dex 
across the two components. The abundances of S and O, both being $\alpha$-elements, track each other. 
Hence, the similarity of the [N/$\alpha$] ratio in both the galaxy and the sub-DLA implies that the 
chemical enrichment history of both are similar. Thus, the absorber may be tracing the extended neutral 
gas disk of the galaxy. Indeed, rotational velocity measurements of the galaxy and the sub-DLA performed 
by \citet{chen2005} hints that that absorbing gas may be corotating along with the optical disk at 
a galactocentric radius of $\sim$ 13.6 kpc (i.e., the deprojected separation of the sub-DLA along
the stellar disk). Alternatively, the gas probed by the sub-DLA could be recently ejected from 
the galaxy and tracing a metal-rich galactic wind/outflow. 

Nitrogen can be of both primary and secondary origin, depending on whether the seed C 
and O are produced by the star itself during helium burning (primary), or whether they 
are from yields of earlier generations of stars and hence already present in the ISM from 
which the star formed (secondary). In nearby galaxies, it has been observed that at lower 
metallicities, i.e., for [O/H] $\le$ 0.4, the [N/O] ratio remains constant (primary N), while 
it rises steeply with increasing O abundance, i.e., for [O/H] $\ge$ 0.7 (secondary N). 
Therefore, the measured [N/$\alpha$] in this sub-DLA along with the high metallicity places 
it in the secondary regime of N production. Hence, the absorbing gas must have undergone a
substantial period of star formation. This reinforces our above hypothesis of either the 
galaxy having an extended neutral disc out to at least $\sim$13 kpc or the sub-DLA tracing 
the metal-enriched halo gas of the galaxy. We test both these scenarios through our photoionization
models in Section~\ref{sec_cloudy}.

Another possible indication that the galaxy might have undergone recent periods of outflow is the presence
of a \lya system at \zabs = 0.10204, $\sim$ $+$250 \kms~from the sub-DLA \citep[see Fig. A2 of][]{muzahid2014}, 
with Ly$\alpha$, $\sit$, $\sitr$ and $\nto$ transitions spread over $\sim$ 200 \kms. We measure log~\nhi = 14.93 $\pm$ 0.44, 
\ln($\sit$) = 13.81 $\pm$ 0.14, \ln($\sitr$) = 13.99 $\pm$ 0.19, and \ln($\nto$) = 14.81 $\pm$ 0.28 
in this system. Based on the large metal column densities, the system could be metal-rich, with 
metallicity and [N/$\alpha$] ratio close to solar. In that case, this cloud is likely to be tracing 
a recent metal-enriched outflow from the galaxy. However, since only the \lya transition is covered 
for this system, the \nhi measurement has large uncertainties and it is plausible that the system has
higher \nhi and lower metallicity.
%
%================================== CLOUDY Discussion =======================================================================
%
\section{Photoionization Models} 
\label{sec_cloudy}
\begin{figure}
\centering \includegraphics[width=0.4\textwidth, angle=90]{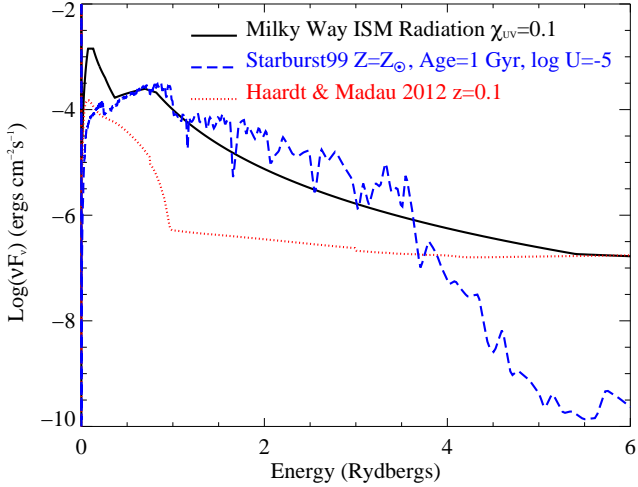}
\caption{Examples of typical incident continuum spectra for different radiation fields used in our photoionization models.}
\label{fig_cont}
\end{figure}

We use the photoionization code \textsc{cloudy} \citep[version 13.03; last described by][]{ferland2013}
to model the physical conditions and infer the chemical enrichment in this system. In all our models, 
the absorber is considered to be a plane-parallel slab of constant density gas with the radiation 
field impinging on it from one side. Note that we do not aim to model all the absorbing ions detected 
from this system due to the multiple-component nature of the absorber and complexities associated with 
phase structure, depletion, and possible hidden saturation in the COS spectrum. We concentrate on \h2, 
$\con$, $\con$*, and $\nao$ absorption arising from Component 2, which are likely to trace the CNM and 
whose excitations will be governed by similar ionizing radiation. The `atom H2' command of \textsc{cloudy}, 
as described in \citet{shaw2005}, is used in order to get an accurate \h2 equilibrium abundance. We consider 
three different incident radiation fields: (i) the metagalactic UV background of QSOs and galaxies \citep[][hereafter HM12]{haardt2012}, 
(ii) the interstellar radiation field as in our Galaxy \citep{black1987}, scaled by a factor \chiu, and 
(iii) a starburst galactic radiation field. Examples of the typical incident continuum spectra for 
different radiation fields used in our models are shown in Fig.~\ref{fig_cont}. The cosmic microwave 
background radiation at $z$ = 0.1 and a cosmic ray ionization rate of log($\Gamma_{CR}$) = $-$17.3 
\citep{williams1998} are also included. For the observed \nhi, we find the ionization correction for 
the S abundance to be negligible (i.e., IC $\le -$ 0.02) compared to the errors, for a wide range of 
density and radiation fields considered. Hence, the average metallicity of the system can be represented 
by [S/H] = 0.28 $\pm$ 0.08, i.e $Z$ $\sim$ 2$Z_{\odot}$. 

In the following sections, we focus our efforts on modelling the strong \h2 component (Component 2) using 
\textsc{cloudy} simulations, that self-consistently compute the ion and molecular abundances along with the 
gas temperature. We run a grid of models by changing the density, \nh~(Sections~\ref{sec_hm12} \& \ref{sec_galrad}), 
or the ionization parameter, $U$ (Section~\ref{sec_stb}), and stopping the calculations when the \Nh2 in the 
model reaches the observed value of 10$^{16.61}$ \cms. Additionally, we discuss the effects of a high \Nh2 
value of 10$^{18.9}$ \cms~(see Section~\ref{sec_h2}) on the models in Section~\ref{sec_highnh2}. We constrain the 
density using the observed limit of $N$($\con$*)/$N$($\con$), the $N$($\con$)/$N$($\nao$) ratio, and by requiring 
the \nhi of the model to be less than the total measured $N$(\hon). Note that, as discussed in Section~\ref{sec_metal}, 
fits to the metal lines in the COS spectrum with more than two components are not preferred. However, fitting 
the $\con$ lines with multiple components having the $z$ and $b$ fixed as that of the $\nao$ lines, allow for 
two times larger $N$($\con$) than that obtained from our best fit. We allow for this uncertainty in the column 
density measurements while comparing with the model predictions. Since there are many parameters in the models, 
for simplicity we take the metallicity in this component to be the same as the average (2$Z_{\odot}$) and the \h2 
formation rate, $R$, to be the same as that given by \citet{jura1975} for our Galaxy. Dust composition is assumed 
similar to the Galaxy and dust depletion is taken as observed for this system. The results of the models are 
summarized in Table~\ref{tab_models}.
\subsection{Models with metagalactic UV background}
\label{sec_hm12}
We first consider the scenario in which the absorber is a gas cloud irradiated with the HM12 background. 
From the total \nhi constraint, we find that the observed \Nh2 can be produced at \nh~$\ge$ 0.03 \cc, 
while from the upper limit of $N$($\con$*)/$N$($\con$), we have \nh~$\le$ 50 \cc. For \nh~$\sim$ 0.1 \cc, 
the $N$($\con$)/$N$($\nao$) ratio, $N$($\con$) and $N$($\nao$) predicted by the model is consistent with 
the observed values within the uncertainties. Moreover, the \th~predicted by this model is consistent with 
our estimation (see Table~\ref{tab_models}). Hence, just the extragalactic background can explain 
the observations provided the gas is at low densities. However, we note that this gas cloud is located just 
outside the optical disk of a star-forming galaxy (see Section~\ref{sec_galaxy}). If we consider an incident 
spectrum similar to that of this galaxy and account for the measured dust extinction (see Section~\ref{sec_stb} 
for details), the expected radiation field near the absorber at 10 eV is $\sim$10 times higher than the HM12 
background. If the $\con$* absorption could be better constrained with higher resolution and S/N data or the 
$\cto$* absorption were covered, it would place additional constraint on the density, and hence allow us to 
test this model better. The fraction of \nhi associated with the \h2 gas in this case is, $f_{\rm N(H\,\textsc{i})}$ $\sim$0.5. 
We estimate the size, $L$, of the \h2 absorbing cloud, assuming spherical geometry, as $\sim$74 pc, using 
$L = f_{\rm N(H\,\textsc{i})} ~ \rm N(H\,\textsc{i}) ~ / ~ ( f_{\rm H\,\textsc{i}} ~ n_{\rm H} )$. \\
\subsection{Models with the mean Galactic radiation field} 
\label{sec_galrad}
Next, we consider the case where the absorber is situated in the extended neutral disc of a galaxy
with radiation field similar to that of the Milky Way. The Galactic radiation field as given in `table ism' 
of \textsc{cloudy} is the unextinguished local interstellar radiation field. However, most of the radiation field 
between 1 to 4 Rydbergs is found to be heavily absorbed by gas in the ISM. Hence, using the `extinguish' 
command of \textsc{cloudy} we introduce photoelectric absorption by a slab of cold neutral gas (\ln($\hon$) $\sim$ 20) 
to mimic typical Galactic ISM sightlines. This is a more appropriate way to model the CNM phase in the gas, 
since this will lead to a single neutral phase throughout the gas slab. Moreover, while the hydrogen ionizing
photons emitted from the galaxy are in principle removed, the species of our interest (\h2, $\con$, $\nao$) 
are susceptible to ionizing photons with energy less than 1 Rydberg. Note that we add the HM12 radiation
field to the Galactic radiation field.

We find that when thermal equilibrium is assumed the \th~estimated by the models is too low ($\sim$30-40 K) compared 
to the observed \th~(133$^{+33}_{-22}$ K). The fact that thermal equilibrium models in \textsc{cloudy} produce very low 
\th, while using the radiation given by the `table ISM' in \textsc{cloudy}, has been noted by \citet{srianand2014} while 
modelling \h2 absorption in a low-$z$ DLA. They suggested that the low temperatures in the models could be due to \textsc{cloudy} 
not considering additional non-radiative heating processes. Additional heating in DLAs can come from cosmic-ray ionization 
\citep[see][]{dutta2014}. However, \citet{srianand2014} have shown that increasing the cosmic-ray ionization rate, while 
increasing the temperature, also causes the ion column densities to increase. Instead, they present \h2 formation at lower 
densities through enhanced formation rate on dust grains as a possible solution for increasing the gas temperature 
\citep[see also][]{habart2011,lebourlot2012}. 

Since the thermal equilibrium models are not able to self-consistently explain the observed temperature, we 
consider constant temperature models, i.e., we fix the temperature to be equal to the observed \th~throughout 
the gas cloud. From these models using the \nhi constraint and the $N$($\con$*)/$N$($\con$) limit, we find the 
\h2 can be produced for \nh~$\sim$ 3$-$30 \cc~and \chiu~$\sim$ 0.1$-$0.5. However, the model-predicted $N$($\con$)/$N$($\nao$) 
ratio for the above solution does not match the observed ratio within the allowed uncertainties (see Table~\ref{tab_models}).
Hence, while the Galactic radiation field attenuated by 10$-$50\% can produce the observed \Nh2, it is unlikely 
to produce the $\con$ and $\nao$ absorption as measured for this gas, and a weaker field may be required. \\

\subsection{Models with starburst radiation field}
\label{sec_stb}
\begin{figure*}
\centering \includegraphics[width=0.6\textwidth, angle=90]{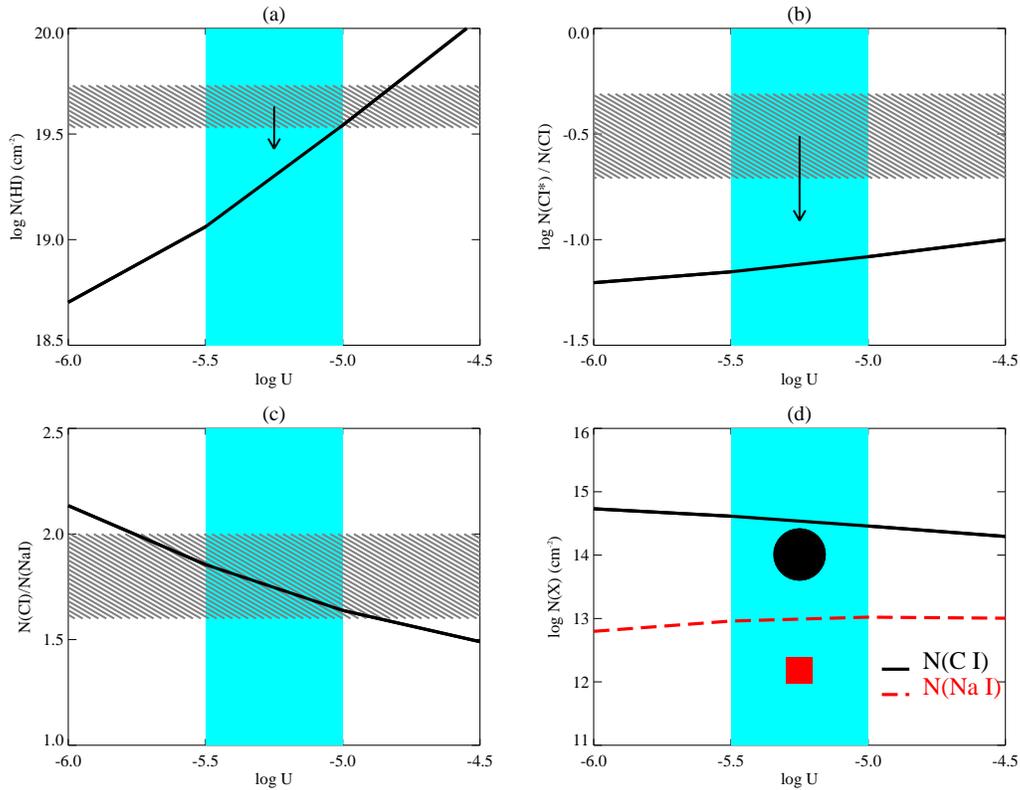}
\caption{Results from photoionization model using the starburst radiation field for the strong \h2 absorbing component 
(Section~\ref{sec_stb}). The model-predicted $N$($\hon$), $N$($\con$*)/$N$($\con$), $N$($\con$)/$N$($\nao$) and 
column densities of $\con$ and $\nao$ as a function of ionization parameter are shown in panels (a), (b), (c) and (d) 
respectively. The vertical shaded regions show the range of $U$ in which the model is consistent with our observations. 
The horizontal shaded regions in (a), (b) and (c) show the allowed range in the values from observations. The symbols 
circle and square in (d) show the measured $N$($\con$) and $N$($\nao$), with the size of the symbols representing the
allowed uncertainties in the measurements.}
\label{fig_models}
\end{figure*}
\noindent Next, we consider the absorber to be present in the halo around a galaxy. \citet{chen2005} have 
measured properties of the galaxy believed to be associated with this sub-DLA (see Section~\ref{sec_galaxy}). 
Using their measurements, we construct a model with radiation field similar to that of the candidate host-galaxy. 
From the H$\alpha$ extinction-corrected luminosity, we estimate a star formation rate of 0.53 M$_{\odot}$yr$^{-1}$ 
\citep{kennicutt1998}. Subsequently, we use \textsc{starburst99}\footnote{http://www.stsci.edu/science/starburst99/docs/table-index.html} 
to generate the spectrum of a continuously star-forming galaxy. The metagalactic HM12 UV background 
is added to the above radiation field. We proceed similarly as above to estimate the physical parameters of 
the system, except in this case we run the model on ionization parameter grids. As in Section~\ref{sec_galrad}, 
the \th~estimated by this model when assuming thermal equilibrium is much lower than our measurement. Hence, we 
consider a model with the temperature kept constant (i.e., equal to \th). We find that such a model is consistent 
with our observations for $-$5.5 $\le$ log~$U$ $\le$ $-$5.0. This can be seen from Table~\ref{tab_models} and 
Fig.~\ref{fig_models}, where we have plotted $N$($\hon$), $N$($\con$*)/$N$($\con$), $N$($\con$)/$N$($\nao$) and 
column densities of $\con$ and $\nao$ obtained from the model as a function of ionization parameter. The $N$($\con$) 
and $N$($\nao$) are over-predicted by the model by 0.3 dex and 0.5 dex respectively, which can be accounted for 
by depletion as seen in the Galactic CNM \citep{welty1999}. Over the range of $U$ where the model is consistent 
with our observations allowing for uncertainties in measurements and depletion effects, $f_{\rm N(H\,\textsc{i})}$ 
is in the range $\sim$ 0.3$-$0.8. 

We define the ionization parameter as, $U = Q \Omega e^{-\tau} /(4 \pi r^{2} n_{H} c)$, where $Q$ is the 
rate at which ionizing photons are emitted by the galaxy, $\Omega$ is the solid angle subtended by the 
galaxy at the absorber, $\tau$ is the dust optical depth, and $r$ is the distance between the galaxy 
and the illuminated face of the cloud. $Q$ can be obtained by integrating the \textsc{starburst99} 
spectrum. Since we are estimating $U$ using $\con$, $\nao$ and \h2 excitations, we estimate the  
dust optical depth at 10 eV ($\tau_{10eV}$). From the extinction measurement of the galaxy, $E(B-V)$ = 
0.22 \citep{chen2005}, the ratio of total-to-selective extinction, $R_{v} = A_{v}/E(B-V)$ = 3.1, and 
the Galactic extinction curve \citep{misselt1999}, we can estimate the extinction expected at 10 eV, 
i.e., $A_{10eV}$ and hence $\tau_{10eV}$. For simplicity, we approximate the galaxy as a circular disc 
of radius 5 kpc with uniform surface brightness, and the absorber as a point at $r$ = 7.6 kpc along 
the normal to the galactic disc \citep{chen2005}, to estimate the solid angle. Then, for the range of 
$U$ obtained from the models, the expected gas density range is, \nh~$\sim$30$-$90 \cc. For the above 
density range and $f_{\rm N(H\,\textsc{i})}$ obtained from the models, the cloud will have size of 
$\sim$ 0.04$-$0.5 pc. We have not taken into account the angle between the absorber and the normal to 
the galactic disc in this simple calculation, which can cause further attenuation in the flux received 
by the absorber, leading to a lower density and a larger cloud size.

Hence, the absorber can be a halo cloud subject to the radiation field due to a continuously star-forming galaxy, 
with metallicity, star formation rate and dust extinction as seen in the candidate host-galaxy. Based on the 
results of our photoionization models and our discussion in Section~\ref{sec_galaxy}, we conclude that the 
absorber is more likely to be tracing gas ejected recently by the galaxy into the circumgalactic medium rather 
than originating in the extended disc. Moreover, we note that in all our photoionization model solutions, the $N$($\aro$) 
predicted by the model is higher (by $\sim$0.5$-$1 dex) than the observed. Argon, being an inert element, is 
not expected to deplete on dust, and we find that $\aro$ is the dominant ionization state of argon in our models. 
As we pointed out before, $\aro$ depletion is seen in the HVCs and in some interstellar sightlines. \citet{sofia1998} 
have argued that this could be related to excess ionization due to hard photons. Our study suggests that for 
the three ionizing radiation fields considered here, Ar depletion cannot be explained by a self-consistent 
ionization model. \citet{jenkins2013} suggested that strong Ar depletion could be due to non-equilibrium 
ionization conditions prevailing in the absorbing gas. Alternatively, the absorbing gas may be tracing a   
non-chemically-well-mixed gas freshly ejected into the halo from a recently exploded supernovae region in 
the galaxy. In order to make headway on this issue it is important to examine several species covering a wide 
rage of ionization states using future COS observations.

\subsection{Models with high \Nh2}
\label{sec_highnh2}
In Section~\ref{sec_h2} we discussed the possibility of a high \Nh2 value (10$^{18.9}$ \cms) for 
Component 2. However, in that case the molecular fraction will be log~f(H$_{2}$) = $-$0.57 if we consider 
all the \nhi to be associated with this component and higher otherwise. For the Galactic disc such high 
molecular fraction generally indicates that the \nhi is above the threshold (10$^{20.7}$ \cms) where the 
\h2 molecule gets completely self-shielded from interstellar radiation \citep{savage1977}. If we assume 
the \h2 formation rate in this system to be same as in our Galaxy, then the high f(H$_{2}$) would require 
either a very low photo-dissociation rate (i.e., a much weaker radiation field) or a very high density. We 
remind here that the observed \th~and fine-structure excitation of C does not allow the density to be more 
than few tens of H atoms \cc~if the relevant parameters are similar to that of the Milky Way (see Section~\ref{sec_carbon}). 
To consider whether such high molecular fraction is feasible in the present system, we ran \textsc{cloudy} 
models with the three background radiation fields considered above and stopped the calculations when log~\Nh2 = 18.9.  
However, the temperature computed in all three cases is much lower (20$-$30 K) than the inferred \th, 
and hence we consider the models with constant temperature. For the extragalactic HM12 background we 
find that the high \Nh2 can be produced for \nh~= 1$-$100 \cc. However, the $N$($\con$) predicted
by this model is more than 1 dex higher than the observed value. For the model with the mean Galactic
radiation field, the \Nh2 can be produced for \nh~= 10$-$30 \cc and \chiu~= 0.1. However, the $N$($\con$)/$N$($\nao$) 
ratio falls $\sim$ 1 dex below the observed value. In case of the starburst radiation field, the
\nhi required by the models to produce the \Nh2 is higher than the total measured \nhi of the system.
Hence, none of the radiation fields considered here is able to consistently explain the high \Nh2 
solution. This supports our argument that the present system is unlikely to host such high molecular 
fraction. However, we emphasize the need for higher S/N and higher resolution data to have better handle 
on the models.
\begin{table*} 
\caption{Summary of the photoionization models for the strong \h2 absorbing component assuming log~\Nh2 = 16.61 as discussed in Section~\ref{sec_cloudy}}
\centering
\begin{tabular}{cccccc}
\hline
\hline
Models $^{a}$ & \nh~(\cc) & $f_{\rm N(H\,\textsc{i})}$  & T (K) & log~$N$($\con$)/$N$($\nao$) & log~$N$($\con$*)/$N$($\con$) \\
\hline
HM12 TE                       & 0.1                     & 0.5         & 160      & 2.0        & $-$2.3                       \\
Galactic \chiu~= 0.1$-$0.5 TE & [1, 100]                & [0.01, 1.0] & [20, 50] & [1.2, 1.5] & [$-$1.2,$-$0.5]              \\ 
Galactic \chiu~= 0.1$-$0.5 CT & [3, 30]                 & [0.1, 1.0]  & 133      & [1.1, 1.2] & [$-$1.3,$-$0.5]              \\
Starburst TE                  & $-$5.5 $^{b}$           & 0.2         & 40       & 1.6        & $-$0.9                       \\
Starburst CT                  & [$-$5.5, $-$5.0] $^{b}$ & [0.3, 0.8]  & 133      & [1.7, 1.8] & [$-$1.1,$-$1.2]              \\
\hline
Observations                  &           &             & 133$^{+33}_{-22}$  & 1.8 $\pm$ 0.2  & $\le$ $-$0.5 \\
\hline
\hline
\end{tabular}
\label{tab_models}
\begin{flushleft}
$^{a}$ TE : thermal equilibrium, CT : constant temperature \\
$^{b}$ log~($U$) \\
\end{flushleft}
\end{table*}
%
%================================== 21-cm Discussion =======================================================================
%
\section{Radio Observations} 
\label{sec_radio}
\subsection{VLBA mas scale imaging}
\label{sec_vlba}
This sub-DLA is unique in the sense that it is the only known low-$z$ (i.e., $z$ $<$ 1.0) 
system with a \h2 detection towards a radio-loud QSO so that \21 absorption observations 
are possible. Given the small inferred sizes from the photoionization models, milliarcsecond 
scale VLBI imaging of the background radio source is necessary to measure the covering factor, 
\fc, of the absorbing gas. The standard practice, in the absence of VLBI spectroscopy, is to 
use the ratio of the VLBI core flux density to the flux density measured in the arcsecond 
scale images to estimate \fc~\citep[see][]{kanekar2009,gupta2012,srianand2012}. 
We carried out continuum observations of the radio source using the VLBA. The spatial 
resolution achieved in our VLBA observations is $\sim$12 mas $\times$ 7 mas, i.e., 
$\sim$23 pc $\times$ 13 pc at the redshift of the absorber. If the size of the 
\h2 gas is of this order, then the fraction and spatial extent of the radio flux 
density detected in the VLBA image will determine the \21 absorption detectability. 

\begin{figure}
\centering \includegraphics[width=0.4\textwidth, angle=270]{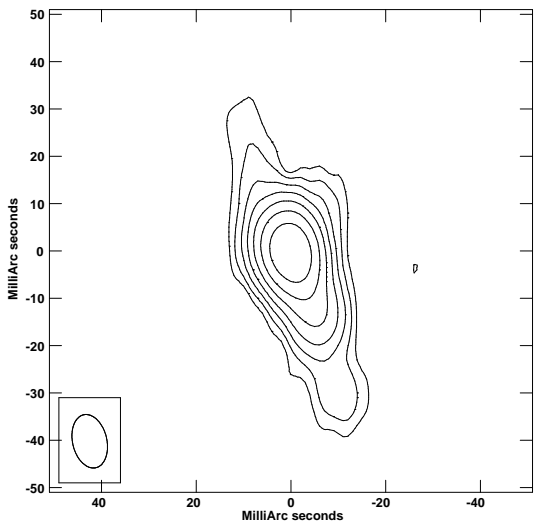}
\caption{Contour plot of the VLBA image of \sys at 1.4 GHz. The rms in the image is 0.15 mJy beam$^{-1}$. 
At the bottom of the image the restoring beam is shown as an ellipse. The beam size is 0.012$''$ $\times$ 0.007$''$. 
The image centre is at RA = 04$^{h}$41$^{m}$17.3367$^{s}$, Dec = $-$43\textdegree13$'$45.4394$''$. 
The contour levels are plotted as 0.46 $\times$ ($-$2,$-$1,1,2,4,8,...) mJy beam$^{-1}$.}
\label{fig_vlba}
\end{figure}

The radio source is identified as a flat-spectrum radio source \citep{healey2007}, 
and is unresolved at arcsecond scales with a flux density of 330 mJy at 1.29 GHz \citep{kanekar2001}. 
However, most of the emission is over-resolved in our VLBA image, with only 27\% being 
recovered, assuming the continuum flux density of the radio source is constant. Our VLBA 
image (Fig.~\ref{fig_vlba}) shows an extended structure with a total flux density of 90 mJy 
and a peak flux density of 59 mJy/beam. Note that we used `ROBUST=2' weighting in the \textsc{aips}
task `IMAGR' to obtain the image shown in Fig.~\ref{fig_vlba}. The largest linear size of 
the radio source measured from the emission detected in the VLBA image is $\sim$70 mas, 
i.e., $\sim$131 pc at the redshift of the absorber. From our discussion on photoionization 
models in the previous section, if the absorbing gas is spherical then the expected extent 
of the strong \h2 absorber is smaller than the extent of the radio emission seen in our 
VLBA image. If we associate the location of the peak emission in the VLBA image to the 
optical source, then the expected covering factor is $\ge$ 0.18. However, if we do not 
assume spherical geometry then we do not have any constraint on the extent of the gas in 
the transverse direction to our line of sight. If we assume the gas to cover all the 
emission seen in the VLBA image, then the covering factor is 0.27. In the following 
section we discussion the implications of this for the detectability of \21 absorption.
\subsection{\21 absorption}
\label{sec_21cm}
In the ATCA radio spectrum (with a channel resolution of 1.8 \kms) reported by \citet{kanekar2001}, 
no \21 absorption is seen at the \zabs of Component 2 at the level of 1.56 mJy (the spectral rms), 
while a line is tentatively detected at 3.3$\sigma$ at \zabs = 0.10097 in a spectrum smoothed to 9 \kms. 
The \21 optical depth $\tau_{21}$ is related to the \nhi (\cms) and \ts (K) as: \nhi = 1.823 $\times$ 10$^{18}$ (\ts$/f_{c}) \int{\tau_{21} dv}$, 
where the $\hon$ \21 line is assumed to be optically thin. In the galactic ISM, the kinetic 
temperature measured from \h2 usually follows the $\hon$ spin temperature \citep{roy2006}. 
We assume this to hold for the present system as well. If all the \nhi were associated with 
Component 2, we would expect a $\sim$3 mJy line, assuming \fc~= 0.27 and a line width of 5 \kms, 
which is ruled out at 2$\sigma$. Hence, the \h2 gas cannot have all the \nhi associated with it 
and cover all the radio emission in the VLBA image. However, if the \h2 gas were to have $\le$ 50\% 
of the total \nhi or cover only part the VLBA emission, the present spectrum would not be sensitive 
enough to detect \21 absorption from this gas. A factor 3 higher S/N spectrum will allow us to detect 
or constrain the \21 absorption at a significance of 3$\sigma$.  

The weak 3.3$\sigma$ \21 detection lies within $\sim$15 \kms~of the weaker \h2 component,
and $\sim$8 \kms~of the weaker components of $\con$, $\cat$ and $\nao$. Recall that 
the measured column densities of metals and dust depletion in this component are not very 
different from that of the strong \h2 component. Usually the absence/weakness of \h2 and $\con$ 
are ascribed to high temperatures and low densities \citep{srianand2005}. In this case we know 
that the temperature cannot be very high, since the \h2, $\con$, $\nao$ absorption are weak,
and the ratio of column densities of the $J$ = 0 \& 1 levels of \h2 in this component suggests
that the temperature is $\ge$ 100 K. If we consider the optical depth reported by \citet{kanekar2001}, 
and assume \fc~= 0.27 and all of the \nhi to be associated with the weak \h2 component, then 
\ts~will be $\sim$90 K. However, from the observed metal content across the components and our 
photoionization models, the weaker component is not likely to have all the \nhi associated with 
it. For $f_{\rm N(H\,I)}$ = 0.5 and \fc~= 0.27, the \ts~will be $\sim$45 K. In case the covering 
factor is lower than 0.27, the inferred \ts~will be even lower. It will be surprising to have such 
a low temperature in a gas with \nhi $\sim$ few 10$^{19}$ \cc. It is most likely that one will need 
higher densities and a lower radiation field. In that case it will be interesting to ask why $\con$ 
lines are weak in such a cold and dense gas. The discussions presented in this section clearly bring
out the importance of having a much higher S/N spectrum to confirm the \21 detection reported by 
\citet{kanekar2001}, and have a stronger constraint on the absence of \21 absorption in the strong 
\h2 component. In addition, better constraints on the $\con$* and $\cto$* absorption will allow us to 
learn more about the physical conditions of the cold gas in this interesting absorber.
%
%=========================== Conclusions & Summary =========================================================================
%
\section{Summary}  
\label{sec_sum} 
We have carried out a detailed analysis of the cold gas phase in the sub-DLA at \zabs = 0.10115 towards J0441$-$4313, 
in which \h2 absorption has been detected by \citet{muzahid2014}. This unique system allows us to study the physical 
conditions in the absorbing gas using both \h2 and \21 absorption simultaneously for the first time at $z$ $<$ 1. 
Below we summarize the main results of our study.
\begin{itemize}
 \item The \h2 absorption arises from one strong component at \zabs = 0.10115, with log~\Nh2 = 
       16.61 $\pm$ 0.05, along with another weaker component at \zabs = 0.10091, with log~\Nh2 $\le$ 
       15.51 $\pm$ 0.03. We note that the \Nh2 measurement of the strong \h2 component is uncertain
       due the medium resolution of the COS spectrum, and that the \Nh2, in principle, can be much 
       higher if the actual $b$ parameter is much smaller. However, the absence of \h2 absorption 
       from higher $J$ levels and of HD absorption, as well as our photionization modelling, do not 
       seem to support a high \Nh2. The excitation temperature (133$^{+33}_{-22}$ K) measured in 
       the strong component for the range log~\Nh2 = 16.6$-$18.9 is similar to what is expected from a CNM phase.
 \item The average metallicity of the absorber is found to be twice solar and the dust depletion 
       moderate ([Fe/S] $\sim$ $-$0.49). We do not find any significant variation in depletion 
       across the two metal components detected in the COS spectrum. The strong \h2 component 
       accounts for 55$-$65\% of the metal column densities. Interestingly, the stronger 
       components of $\con$, $\nao$ and $\aro$ are coincident with this component. 
 \item The sub-DLA is known to be associated with a star-forming galaxy at a projected separation 
       of $\sim$ 7.6 kpc, just outside the optical galactic disk \citep{chen2005}. We do not find any 
       metallicity gradient between the host-galaxy and the sub-DLA as suggested by \citet{chen2005}.
       Moreover, the [N/$\alpha$] ratio measured in both are similar. Hence, the galaxy is likely to 
       either have a metal-rich neutral disk extending to at least $\sim$13 kpc (the galactocentric 
       radius of the sub-DLA) or have undergone recent periods of metal-rich outflow.
 \item From photoionization modelling, we show that the observed column densities of \h2, $\con$ and $\nao$ in 
       the \zabs~= 0.10115 component can be consistently explained using a radiation field due to a continuously 
       star-forming galaxy, with metallicity, star formation rate and dust extinction as measured in the associated 
       galaxy. Alternatively, if the absorber is tracing the extended galactic disc then the radiation field has to 
       be weaker than half the mean Galactic radiation field. However, the models suggest that the absorber is more 
       likely to be tracing gas in the galactic halo than in the extended galactic disc. We note that, the measured 
       column densities are also consistent with the extragalactic background radiation if the gas has \nh~$\sim$ 
       0.1 \cc~and $L$ $\sim$70 pc. However, from the presence of a star-forming galaxy at $\sim$7.6 kpc, we 
       argue that this absorber is more likely to be a halo cloud. In that case, using simple approximations we 
       obtain for this gas cloud, \nh~$\sim$ 30$-$90 \cc~and $L$ $\sim$ 0.05$-$0.4 pc.
 \item The measured $\aro$ absorption strength in this system is much weaker than what is expected if there is 
       no depletion or ionization effects, and than what is predicted by all our photoionization models. Similar
       depletion is seen in case of HVCs in our Galaxy. The possible explanations could be non-equilibrium
       ionization conditions in the absorber or the gas not being chemically well-mixed due to recent ejection
       from supernovae region in the associated galaxy.
 \item This sub-DLA presents a fortuitous case of a \h2 absorber towards a radio-loud QSO which can be searched 
       for \21 absorption. We present the VLBA mas scale image of the background radio source, which shows it to 
       be over-resolved, with only 27\% of the arcsecond flux being recovered. The radio emission has a total 
       flux density of 90 mJy, a peak flux density of 59 mJy/beam and an extent of $\sim$131 pc at the redshift 
       of the absorber.   
 \item \citet{kanekar2001} report a tentative detection of weak \21 absorption that arises within $\sim$15 \kms~of 
       the weaker \h2 component. However, \21 absorption from the stronger \h2 component is absent at the level of 
       2$\sigma$, indicating that either the \nhi associated with this component is $\le$ 50\% of the total measured 
       \nhi or that the covering factor of this gas is $\le$ 0.27, consistent with results from our photoionization models. 
 \item The reported \21 absorption from the weaker \h2 component indicates that \ts~$\le$ 90 K in this gas. The actual
       value could be much lower if a good fraction of the observed \nhi is associated with the strong \h2 component.
       However, the weakness of \h2, $\con$ and $\nao$ absorption from such cold gas is puzzling. A higher S/N \21 
       absorption spectrum and a UV spectrum with better coverage of species like $\con$, $\con$* and $\cto$* are 
       essential to put more stringent constraints on the conditions in this absorbing gas.
\end{itemize} 
Such detailed analysis of similar individual systems is essential to further our understanding of the properties 
of cold gas present around galaxies, and in particular, of the connection between \h2 and \21 absorption. The results
presented here will also facilitate interpretation of the results of large surveys of \h2 and \21 absorption. \\    
%
%========================================= Acknowledgment ==================================================================
%

\noindent \textbf{ACKNOWLEDGEMENTS} \newline

\noindent We thank the anonymous referee for his/her useful comments. The VLBA is run by the National Radio Astronomy Observatory. 
The National Radio Astronomy Observatory is a facility of the National Science Foundation operated under cooperative agreement by 
Associated Universities, Inc. This work made use of the Swinburne University of Technology software correlator, developed as part 
of the Australian Major National Research Facilities Program and operated under license \citep{deller2011}.
%
%======================================== Bibliography =====================================================================
% 
\def\aj{AJ}%
\def\actaa{Acta Astron.}%
\def\araa{ARA\&A}%
\def\apj{ApJ}%
\def\apjl{ApJ}%
\def\apjs{ApJS}%
\def\ao{Appl.~Opt.}%
\def\apss{Ap\&SS}%
\def\aap{A\&A}%
\def\aapr{A\&A~Rev.}%
\def\aaps{A\&AS}%
\def\azh{A$Z$h}%
\def\baas{BAAS}%
\def\bac{Bull. astr. Inst. Czechosl.}%
\def\caa{Chinese Astron. Astrophys.}%
\def\cjaa{Chinese J. Astron. Astrophys.}%
\def\icarus{Icarus}%
\def\jcap{J. Cosmology Astropart. Phys.}%
\def\jrasc{JRASC}%
\def\mnras{MNRAS}%
\def\memras{MmRAS}%
\def\na{New A}%
\def\nar{New A Rev.}%
\def\pasa{PASA}%
\def\pra{Phys.~Rev.~A}%
\def\prb{Phys.~Rev.~B}%
\def\prc{Phys.~Rev.~C}%
\def\prd{Phys.~Rev.~D}%
\def\pre{Phys.~Rev.~E}%
\def\prl{Phys.~Rev.~Lett.}%
\def\pasp{PASP}%
\def\pasj{PASJ}%
\def\qjras{QJRAS}%
\def\rmxaa{Rev. Mexicana Astron. Astrofis.}%
\def\skytel{S\&T}%
\def\solphys{Sol.~Phys.}%
\def\sovast{Soviet~Ast.}%
\def\ssr{Space~Sci.~Rev.}%
\def\zap{$Z$Ap}%
\def\nat{Nature}%
\def\iaucirc{IAU~Circ.}%
\def\aplett{Astrophys.~Lett.}%
\def\apspr{Astrophys.~Space~Phys.~Res.}%
\def\bain{Bull.~Astron.~Inst.~Netherlands}%
\def\fcp{Fund.~Cosmic~Phys.}%
\def\gca{Geochim.~Cosmochim.~Acta}%
\def\grl{Geophys.~Res.~Lett.}%
\def\jcp{J.~Chem.~Phys.}%
\def\jgr{J.~Geophys.~Res.}%
\def\jqsrt{J.~Quant.~Spec.~Radiat.~Transf.}%
\def\memsai{Mem.~Soc.~Astron.~Italiana}%
\def\nphysa{Nucl.~Phys.~A}%
\def\physrep{Phys.~Rep.}%
\def\physscr{Phys.~Scr}%
\def\planss{Planet.~Space~Sci.}%
\def\procspie{Proc.~SPIE}%
\let\astap=\aap
\let\apjlett=\apjl
\let\apjsupp=\apjs
\let\applopt=\ao
\bibliographystyle{mn}
\bibliography{mybib}

\begin{thebibliography}{75}
\expandafter\ifx\csname natexlab\endcsname\relax\def\natexlab#1{#1}\fi

\bibitem[{{Akaike}(1974)}]{akaike1974}
{Akaike}, H., 1974, IEEE Transactions on Automatic Control, 19, 716

\bibitem[{{Asplund} {et~al.}(2009){Asplund}, {Grevesse}, {Sauval}, \&
  {Scott}}]{asplund2009}
{Asplund}, M., {Grevesse}, N., {Sauval}, A.~J., \& {Scott}, P., 2009, \araa,
  47, 481

\bibitem[{{Balashev} {et~al.}(2011){Balashev}, {Petitjean}, {Ivanchik},
  {Ledoux}, {Srianand}, {Noterdaeme}, \& {Varshalovich}}]{balashev2011}
{Balashev}, S.~A., {Petitjean}, P., {Ivanchik}, A.~V., {Ledoux}, C.,
  {Srianand}, R., {Noterdaeme}, P., \& {Varshalovich}, D.~A., 2011, \mnras,
  418, 357

\bibitem[{{Battisti} {et~al.}(2012){Battisti}, {Meiring}, {Tripp}, {Prochaska},
  {Werk}, {Jenkins}, {Lehner}, {Tumlinson}, \& {Thom}}]{battisti2012}
{Battisti}, A.~J., {Meiring}, J.~D., {Tripp}, T.~M., {et~al.}, 2012, \apj, 744,
  93

\bibitem[{{Black} \& {van Dishoeck}(1987)}]{black1987}
{Black}, J.~H. \& {van Dishoeck}, E.~F., 1987, \apj, 322, 412

\bibitem[{{Borthakur} {et~al.}(2010){Borthakur}, {Tripp}, {Yun}, {Momjian},
  {Meiring}, {Bowen}, \& {York}}]{borthakur2010}
{Borthakur}, S., {Tripp}, T.~M., {Yun}, M.~S., {Momjian}, E., {Meiring}, J.~D.,
  {Bowen}, D.~V., \& {York}, D.~G., 2010, \apj, 713, 131

\bibitem[{{Chen} {et~al.}(2005){Chen}, {Kennicutt}, \& {Rauch}}]{chen2005}
{Chen}, H.-W., {Kennicutt}, Jr., R.~C., \& {Rauch}, M., 2005, \apj, 620, 703

\bibitem[{{Churchill}(2001)}]{churchill2001}
{Churchill}, C.~W., 2001, \apj, 560, 92

\bibitem[{{Crighton} {et~al.}(2013){Crighton}, {Bechtold}, {Carswell},
  {Dav{\'e}}, {Foltz}, {Jannuzi}, {Morris}, {O'Meara}, {Prochaska}, {Schaye},
  \& {Tejos}}]{crighton2013}
{Crighton}, N.~H.~M., {Bechtold}, J., {Carswell}, R.~F., {et~al.}, 2013,
  \mnras, 433, 178

\bibitem[{{Curran} {et~al.}(2005){Curran}, {Murphy}, {Pihlstr{\"o}m}, {Webb},
  \& {Purcell}}]{curran2005}
{Curran}, S.~J., {Murphy}, M.~T., {Pihlstr{\"o}m}, Y.~M., {Webb}, J.~K., \&
  {Purcell}, C.~R., 2005, \mnras, 356, 1509

\bibitem[{{Danforth} {et~al.}(2010){Danforth}, {Keeney}, {Stocke}, {Shull}, \&
  {Yao}}]{danforth2010}
{Danforth}, C.~W., {Keeney}, B.~A., {Stocke}, J.~T., {Shull}, J.~M., \& {Yao},
  Y., 2010, \apj, 720, 976

\bibitem[{{Deller} {et~al.}(2011){Deller}, {Brisken}, {Phillips}, {Morgan},
  {Alef}, {Cappallo}, {Middelberg}, {Romney}, {Rottmann}, {Tingay}, \&
  {Wayth}}]{deller2011}
{Deller}, A.~T., {Brisken}, W.~F., {Phillips}, C.~J., {et~al.}, 2011, \pasp,
  123, 275

\bibitem[{{Dutta} {et~al.}(2014){Dutta}, {Srianand}, {Rahmani}, {Petitjean},
  {Noterdaeme}, \& {Ledoux}}]{dutta2014}
{Dutta}, R., {Srianand}, R., {Rahmani}, H., {Petitjean}, P., {Noterdaeme}, P.,
  \& {Ledoux}, C., 2014, \mnras, 440, 307

\bibitem[{{Edlen}(1966)}]{edlen1966}
{Edlen}, B., 1966, Transactions of the International Astronomical Union, Series
  B, 12, 176

\bibitem[{{Ferland} {et~al.}(2013){Ferland}, {Porter}, {van Hoof}, {Williams},
  {Abel}, {Lykins}, {Shaw}, {Henney}, \& {Stancil}}]{ferland2013}
{Ferland}, G.~J., {Porter}, R.~L., {van Hoof}, P.~A.~M., {et~al.}, 2013,
  \rmxaa, 49, 137

\bibitem[{{Ferlet} {et~al.}(1985){Ferlet}, {Vidal-Madjar}, \&
  {Gry}}]{ferlet1985}
{Ferlet}, R., {Vidal-Madjar}, A., \& {Gry}, C., 1985, \apj, 298, 838

\bibitem[{{Ghavamian} {et~al.}(2009){Ghavamian}, {Aloisi}, {Lennon}, {Hartig},
  {Kriss}, {Oliveira}, {Massa}, {Keyes}, {Proffitt}, {Delker}, \&
  {Osterman}}]{ghavamian2009}
{Ghavamian}, P., {Aloisi}, A., {Lennon}, D., {et~al.}, 2009, {Preliminary
  Characterization of the Post- Launch Line Spread Function of COS}. Tech. rep.

\bibitem[{{Greisen}(2003)}]{greisen2003}
{Greisen}, E.~W., 2003, Information Handling in Astronomy - Historical Vistas,
  285, 109

\bibitem[{{Gupta} {et~al.}(2012){Gupta}, {Srianand}, {Petitjean}, {Bergeron},
  {Noterdaeme}, \& {Muzahid}}]{gupta2012}
{Gupta}, N., {Srianand}, R., {Petitjean}, P., {Bergeron}, J., {Noterdaeme}, P.,
  \& {Muzahid}, S., 2012, \aap, 544, A21

\bibitem[{{Haardt} \& {Madau}(2012)}]{haardt2012}
{Haardt}, F. \& {Madau}, P., 2012, \apj, 746, 125

\bibitem[{{Habart} {et~al.}(2011){Habart}, {Abergel}, {Boulanger}, {Joblin},
  {Verstraete}, {Compi{\`e}gne}, {Pineau Des For{\^e}ts}, \& {Le
  Bourlot}}]{habart2011}
{Habart}, E., {Abergel}, A., {Boulanger}, F., {Joblin}, C., {Verstraete}, L.,
  {Compi{\`e}gne}, M., {Pineau Des For{\^e}ts}, G., \& {Le Bourlot}, J., 2011,
  \aap, 527, A122

\bibitem[{{Healey} {et~al.}(2007){Healey}, {Romani}, {Taylor}, {Sadler},
  {Ricci}, {Murphy}, {Ulvestad}, \& {Winn}}]{healey2007}
{Healey}, S.~E., {Romani}, R.~W., {Taylor}, G.~B., {Sadler}, E.~M., {Ricci},
  R., {Murphy}, T., {Ulvestad}, J.~S., \& {Winn}, J.~N., 2007, \apjs, 171, 61

\bibitem[{{Heiles} \& {Troland}(2004)}]{heiles2004}
{Heiles}, C. \& {Troland}, T.~H., 2004, \apjs, 151, 271

\bibitem[{{Izotov} {et~al.}(2006){Izotov}, {Stasi{\'n}ska}, {Meynet}, {Guseva},
  \& {Thuan}}]{izotov2006}
{Izotov}, Y.~I., {Stasi{\'n}ska}, G., {Meynet}, G., {Guseva}, N.~G., \&
  {Thuan}, T.~X., 2006, \aap, 448, 955

\bibitem[{{Jenkins}(2013)}]{jenkins2013}
{Jenkins}, E.~B., 2013, \apj, 764, 25

\bibitem[{{Jura}(1975)}]{jura1975}
{Jura}, M., 1975, \apj, 197, 575

\bibitem[{{Kanekar} {et~al.}(2001){Kanekar}, {Chengalur}, {Subrahmanyan}, \&
  {Petitjean}}]{kanekar2001}
{Kanekar}, N., {Chengalur}, J.~N., {Subrahmanyan}, R., \& {Petitjean}, P.,
  2001, \aap, 367, 46

\bibitem[{{Kanekar} {et~al.}(2009){Kanekar}, {Lane}, {Momjian}, {Briggs}, \&
  {Chengalur}}]{kanekar2009}
{Kanekar}, N., {Lane}, W.~M., {Momjian}, E., {Briggs}, F.~H., \& {Chengalur},
  J.~N., 2009, \mnras, 394, L61

\bibitem[{{Kanekar} {et~al.}(2014){Kanekar}, {Prochaska}, {Smette}, {Ellison},
  {Ryan-Weber}, {Momjian}, {Briggs}, {Lane}, {Chengalur}, {Delafosse}, {Grave},
  {Jacobsen}, \& {de Bruyn}}]{kanekar2014}
{Kanekar}, N., {Prochaska}, J.~X., {Smette}, A., {et~al.}, 2014, \mnras, 438,
  2131

\bibitem[{{Kennicutt}(1998)}]{kennicutt1998}
{Kennicutt}, Jr., R.~C., 1998, \apj, 498, 541

\bibitem[{{King}(2011)}]{king2011}
{King}, J., 2011, ArXiv e-prints

\bibitem[{{Kriss}(2011)}]{kriss2011}
{Kriss}, G.~A., 2011, {Improved Medium Resolution Line Spread Functions for COS
  FUV Spectra}. Tech. rep.

\bibitem[{{Kulkarni} \& {Heiles}(1988)}]{kulkarni1988}
{Kulkarni}, S.~R. \& {Heiles}, C., 1988, {Neutral hydrogen and the diffuse
  interstellar medium}, {Kellermann}, K.~I. \& {Verschuur}, G.~L., eds., pp.
  95--153

\bibitem[{{Le Bourlot} {et~al.}(2012){Le Bourlot}, {Le Petit}, {Pinto},
  {Roueff}, \& {Roy}}]{lebourlot2012}
{Le Bourlot}, J., {Le Petit}, F., {Pinto}, C., {Roueff}, E., \& {Roy}, F.,
  2012, \aap, 541, A76

\bibitem[{{Ledoux} {et~al.}(2003){Ledoux}, {Petitjean}, \&
  {Srianand}}]{ledoux2003}
{Ledoux}, C., {Petitjean}, P., \& {Srianand}, R., 2003, \mnras, 346, 209

\bibitem[{{Meiring} {et~al.}(2011){Meiring}, {Tripp}, {Prochaska}, {Tumlinson},
  {Werk}, {Jenkins}, {Thom}, {O'Meara}, \& {Sembach}}]{meiring2011}
{Meiring}, J.~D., {Tripp}, T.~M., {Prochaska}, J.~X., {et~al.}, 2011, \apj,
  732, 35

\bibitem[{{Meiring} {et~al.}(2013){Meiring}, {Tripp}, {Werk}, {Howk},
  {Jenkins}, {Prochaska}, {Lehner}, \& {Sembach}}]{meiring2013}
{Meiring}, J.~D., {Tripp}, T.~M., {Werk}, J.~K., {Howk}, J.~C., {Jenkins},
  E.~B., {Prochaska}, J.~X., {Lehner}, N., \& {Sembach}, K.~R., 2013, \apj,
  767, 49

\bibitem[{{Misselt} {et~al.}(1999){Misselt}, {Clayton}, \&
  {Gordon}}]{misselt1999}
{Misselt}, K.~A., {Clayton}, G.~C., \& {Gordon}, K.~D., 1999, \apj, 515, 128

\bibitem[{{Momjian} {et~al.}(2002){Momjian}, {Romney}, \&
  {Troland}}]{momjian2002}
{Momjian}, E., {Romney}, J.~D., \& {Troland}, T.~H., 2002, \apj, 566, 195

\bibitem[{{Muzahid} {et~al.}(2014){Muzahid}, {Srianand}, \&
  {Charlton}}]{muzahid2014}
{Muzahid}, S., {Srianand}, R., \& {Charlton}, J., 2014, ArXiv eprints 1410.3828

\bibitem[{{Noterdaeme} {et~al.}(2008{\natexlab{a}}){Noterdaeme}, {Ledoux},
  {Petitjean}, \& {Srianand}}]{noterdaeme2008a}
{Noterdaeme}, P., {Ledoux}, C., {Petitjean}, P., \& {Srianand}, R.,
  2008{\natexlab{a}}, \aap, 481, 327

\bibitem[{{Noterdaeme} {et~al.}(2009{\natexlab{a}}){Noterdaeme}, {Ledoux},
  {Srianand}, {Petitjean}, \& {Lopez}}]{noterdaeme2009a}
{Noterdaeme}, P., {Ledoux}, C., {Srianand}, R., {Petitjean}, P., \& {Lopez},
  S., 2009{\natexlab{a}}, \aap, 503, 765

\bibitem[{{Noterdaeme} {et~al.}(2012){Noterdaeme}, {Petitjean}, {Carithers},
  {P{\^a}ris}, {Font-Ribera}, {Bailey}, {Aubourg}, {Bizyaev}, {Ebelke},
  {Finley}, {Ge}, {Malanushenko}, {Malanushenko}, {Miralda-Escud{\'e}},
  {Myers}, {Oravetz}, {Pan}, {Pieri}, {Ross}, {Schneider}, {Simmons}, \&
  {York}}]{noterdaeme2012}
{Noterdaeme}, P., {Petitjean}, P., {Carithers}, W.~C., {et~al.}, 2012, \aap,
  547, L1

\bibitem[{{Noterdaeme} {et~al.}(2009{\natexlab{b}}){Noterdaeme}, {Petitjean},
  {Ledoux}, \& {Srianand}}]{noterdaeme2009b}
{Noterdaeme}, P., {Petitjean}, P., {Ledoux}, C., \& {Srianand}, R.,
  2009{\natexlab{b}}, \aap, 505, 1087

\bibitem[{{Noterdaeme} {et~al.}(2008{\natexlab{b}}){Noterdaeme}, {Petitjean},
  {Ledoux}, {Srianand}, \& {Ivanchik}}]{noterdaeme2008b}
{Noterdaeme}, P., {Petitjean}, P., {Ledoux}, C., {Srianand}, R., \& {Ivanchik},
  A., 2008{\natexlab{b}}, \aap, 491, 397

\bibitem[{{Noterdaeme} {et~al.}(2007){Noterdaeme}, {Petitjean}, {Srianand},
  {Ledoux}, \& {Le Petit}}]{noterdaeme2007}
{Noterdaeme}, P., {Petitjean}, P., {Srianand}, R., {Ledoux}, C., \& {Le Petit},
  F., 2007, \aap, 469, 425

\bibitem[{{Oliveira} {et~al.}(2014){Oliveira}, {Sembach}, {Tumlinson},
  {O'Meara}, \& {Thom}}]{oliveira2014}
{Oliveira}, C.~M., {Sembach}, K.~R., {Tumlinson}, J., {O'Meara}, J., \& {Thom},
  C., 2014, \apj, 783, 22

\bibitem[{{Pequignot} \& {Aldrovandi}(1986)}]{pequignot1986}
{Pequignot}, D. \& {Aldrovandi}, S.~M.~V., 1986, \aap, 161, 169

\bibitem[{{P{\'e}roux} {et~al.}(2012){P{\'e}roux}, {Bouch{\'e}}, {Kulkarni},
  {York}, \& {Vladilo}}]{peroux2012}
{P{\'e}roux}, C., {Bouch{\'e}}, N., {Kulkarni}, V.~P., {York}, D.~G., \&
  {Vladilo}, G., 2012, \mnras, 419, 3060

\bibitem[{{P{\'e}roux} {et~al.}(2005){P{\'e}roux}, {Dessauges-Zavadsky},
  {D'Odorico}, {Sun Kim}, \& {McMahon}}]{peroux2005}
{P{\'e}roux}, C., {Dessauges-Zavadsky}, M., {D'Odorico}, S., {Sun Kim}, T., \&
  {McMahon}, R.~G., 2005, \mnras, 363, 479

\bibitem[{{Petitjean} {et~al.}(2006){Petitjean}, {Ledoux}, {Noterdaeme}, \&
  {Srianand}}]{petitjean2006}
{Petitjean}, P., {Ledoux}, C., {Noterdaeme}, P., \& {Srianand}, R., 2006, \aap,
  456, L9

\bibitem[{{Petitjean} {et~al.}(1996){Petitjean}, {Theodore}, {Smette}, \&
  {Lespine}}]{petitjean1996}
{Petitjean}, P., {Theodore}, B., {Smette}, A., \& {Lespine}, Y., 1996, \aap,
  313, L25

\bibitem[{{Rao} {et~al.}(2011){Rao}, {Belfort-Mihalyi}, {Turnshek}, {Monier},
  {Nestor}, \& {Quider}}]{rao2011}
{Rao}, S.~M., {Belfort-Mihalyi}, M., {Turnshek}, D.~A., {Monier}, E.~M.,
  {Nestor}, D.~B., \& {Quider}, A., 2011, \mnras, 416, 1215

\bibitem[{{Richter} {et~al.}(2011){Richter}, {Krause}, {Fechner}, {Charlton},
  \& {Murphy}}]{richter2011}
{Richter}, P., {Krause}, F., {Fechner}, C., {Charlton}, J.~C., \& {Murphy},
  M.~T., 2011, \aap, 528, A12

\bibitem[{{Richter} {et~al.}(2001){Richter}, {Sembach}, {Wakker}, {Savage},
  {Tripp}, {Murphy}, {Kalberla}, \& {Jenkins}}]{richter2001}
{Richter}, P., {Sembach}, K.~R., {Wakker}, B.~P., {Savage}, B.~D., {Tripp},
  T.~M., {Murphy}, E.~M., {Kalberla}, P.~M.~W., \& {Jenkins}, E.~B., 2001,
  \apj, 559, 318

\bibitem[{{Roy} {et~al.}(2006){Roy}, {Chengalur}, \& {Srianand}}]{roy2006}
{Roy}, N., {Chengalur}, J.~N., \& {Srianand}, R., 2006, \mnras, 365, L1

\bibitem[{{Savage} {et~al.}(1977){Savage}, {Bohlin}, {Drake}, \&
  {Budich}}]{savage1977}
{Savage}, B.~D., {Bohlin}, R.~C., {Drake}, J.~F., \& {Budich}, W., 1977, \apj,
  216, 291

\bibitem[{{Savage} {et~al.}(2011){Savage}, {Narayanan}, {Lehner}, \&
  {Wakker}}]{savage2011}
{Savage}, B.~D., {Narayanan}, A., {Lehner}, N., \& {Wakker}, B.~P., 2011, \apj,
  731, 14

\bibitem[{{Shaw} {et~al.}(2005){Shaw}, {Ferland}, {Abel}, {Stancil}, \& {van
  Hoof}}]{shaw2005}
{Shaw}, G., {Ferland}, G.~J., {Abel}, N.~P., {Stancil}, P.~C., \& {van Hoof},
  P.~A.~M., 2005, \apj, 624, 794

\bibitem[{{Sofia} \& {Jenkins}(1998)}]{sofia1998}
{Sofia}, U.~J. \& {Jenkins}, E.~B., 1998, \apj, 499, 951

\bibitem[{{Srianand} {et~al.}(2012){Srianand}, {Gupta}, {Petitjean},
  {Noterdaeme}, {Ledoux}, {Salter}, \& {Saikia}}]{srianand2012}
{Srianand}, R., {Gupta}, N., {Petitjean}, P., {Noterdaeme}, P., {Ledoux}, C.,
  {Salter}, C.~J., \& {Saikia}, D.~J., 2012, \mnras, 421, 651

\bibitem[{{Srianand} {et~al.}(2013){Srianand}, {Gupta}, {Rahmani}, {Momjian},
  {Petitjean}, \& {Noterdaeme}}]{srianand2013}
{Srianand}, R., {Gupta}, N., {Rahmani}, H., {Momjian}, E., {Petitjean}, P., \&
  {Noterdaeme}, P., 2013, \mnras, 428, 2198

\bibitem[{{Srianand} {et~al.}(2008){Srianand}, {Noterdaeme}, {Ledoux}, \&
  {Petitjean}}]{srianand2008}
{Srianand}, R., {Noterdaeme}, P., {Ledoux}, C., \& {Petitjean}, P., 2008, \aap,
  482, L39

\bibitem[{{Srianand} {et~al.}(2005){Srianand}, {Petitjean}, {Ledoux},
  {Ferland}, \& {Shaw}}]{srianand2005}
{Srianand}, R., {Petitjean}, P., {Ledoux}, C., {Ferland}, G., \& {Shaw}, G.,
  2005, \mnras, 362, 549

\bibitem[{{Srianand} {et~al.}(2014){Srianand}, {Rahmani}, {Muzahid}, \&
  {Mohan}}]{srianand2014}
{Srianand}, R., {Rahmani}, H., {Muzahid}, S., \& {Mohan}, V., 2014, \mnras,
  443, 3318

\bibitem[{{Sugiura}(1978)}]{sugiura1978}
{Sugiura}, N., 1978, Communications in Statistics - Theory and Methods, A7, 13

\bibitem[{{Varshalovich} {et~al.}(2001){Varshalovich}, {Ivanchik}, {Petitjean},
  {Srianand}, \& {Ledoux}}]{varshalovich2001}
{Varshalovich}, D.~A., {Ivanchik}, A.~V., {Petitjean}, P., {Srianand}, R., \&
  {Ledoux}, C., 2001, Astronomy Letters, 27, 683

\bibitem[{{Wakker} \& {Mathis}(2000)}]{wakker2000}
{Wakker}, B.~P. \& {Mathis}, J.~S., 2000, \apjl, 544, L107

\bibitem[{{Welty} {et~al.}(1999){Welty}, {Hobbs}, {Lauroesch}, {Morton},
  {Spitzer}, \& {York}}]{welty1999}
{Welty}, D.~E., {Hobbs}, L.~M., {Lauroesch}, J.~T., {Morton}, D.~C., {Spitzer},
  L., \& {York}, D.~G., 1999, \apjs, 124, 465

\bibitem[{{Welty} {et~al.}(1996){Welty}, {Morton}, \& {Hobbs}}]{welty1996}
{Welty}, D.~E., {Morton}, D.~C., \& {Hobbs}, L.~M., 1996, \apjs, 106, 533

\bibitem[{{Williams} {et~al.}(1998){Williams}, {Bergin}, {Caselli}, {Myers}, \&
  {Plume}}]{williams1998}
{Williams}, J.~P., {Bergin}, E.~A., {Caselli}, P., {Myers}, P.~C., \& {Plume},
  R., 1998, \apj, 503, 689

\bibitem[{{Wolfe} {et~al.}(2005){Wolfe}, {Gawiser}, \& {Prochaska}}]{wolfe2005}
{Wolfe}, A.~M., {Gawiser}, E., \& {Prochaska}, J.~X., 2005, \araa, 43, 861

\bibitem[{{Wolfe} {et~al.}(2003){Wolfe}, {Prochaska}, \& {Gawiser}}]{wolfe2003}
{Wolfe}, A.~M., {Prochaska}, J.~X., \& {Gawiser}, E., 2003, \apj, 593, 215

\bibitem[{{Wolfire} {et~al.}(1995){Wolfire}, {Hollenbach}, {McKee}, {Tielens},
  \& {Bakes}}]{wolfire1995}
{Wolfire}, M.~G., {Hollenbach}, D., {McKee}, C.~F., {Tielens}, A.~G.~G.~M., \&
  {Bakes}, E.~L.~O., 1995, \apj, 443, 152

\bibitem[{{Zafar} {et~al.}(2014){Zafar}, {Vladilo}, {P{\'e}roux}, {Molaro},
  {Centuri{\'o}n}, {D'Odorico}, {Abbas}, \& {Popping}}]{zafar2014}
{Zafar}, T., {Vladilo}, G., {P{\'e}roux}, C., {Molaro}, P., {Centuri{\'o}n},
  M., {D'Odorico}, V., {Abbas}, K., \& {Popping}, A., 2014, \mnras, 445, 2093

\end{thebibliography}
\end{document}